\documentclass[conference,compsoc]{IEEEtran}

\newif\iffull
\fulltrue
\newif\ifdraft
\draftfalse

\usepackage[utf8]{inputenc}
\usepackage{amssymb}
\usepackage{amsmath}
\usepackage{stmaryrd}
\usepackage{amsfonts}
\usepackage{mathtools}
\usepackage{xspace}
\usepackage{infer}
\usepackage{float}
\usepackage[all]{xy}
\usepackage{array}
\usepackage{version}
\usepackage{balance}
\usepackage{enumerate}
\usepackage{url}
\usepackage{multirow}
\usepackage{xspace}

\usepackage[users=IMS,\ifdraft\else hide\fi]{uchanges}
\ChangesSetUserColor{I}{green}
\ChangesSetUserColor{M}{blue}
\ChangesSetUserColor{S}{red}

\floatstyle{ruled}
\restylefloat{table}

\begin{document}

\title{\tool: Practical and Sound Static Analysis \\ of Android Applications by
SMT Solving}

\author{\IEEEauthorblockN{Stefano Calzavara}
\IEEEauthorblockA{Universit\`{a} Ca' Foscari Venezia \\
calzavara@dais.unive.it} 

\and

\IEEEauthorblockN{Ilya Grishchenko}
\IEEEauthorblockA{CISPA, Saarland University \\
grishchenko@cs.uni-saarland.de}

\and

\IEEEauthorblockN{Matteo Maffei}
\IEEEauthorblockA{CISPA, Saarland University \\
maffei@cs.uni-saarland.de}}

\newtheorem{definition}{Definition}
\newtheorem{lemma}{Lemma}
\newtheorem{assumption}{Assumption}
\newtheorem{theorem}{Theorem}
\newcommand{\dom}{\textit{dom}}
\newcommand{\irule}[1]{({\sc #1})}
\newcommand{\sem}{{$\mu\text{-Dalvik}_{A}$}\xspace}
\newcommand{\etal}{\textit{et al.}}
\newcommand{\tool}{{HornDroid}\xspace}

\newcommand{\yt}{\emph{yes}}
\newcommand{\yf}{yes}
\newcommand{\nt}{\emph{no}}
\newcommand{\nf}{no}

\newcommand{\class}{\mathit{cls}}
\newcommand{\cls}[5]{\texttt{cls}\ #1 \subtype #2\ \texttt{imp}\ #3\ \{#4; #5\}}
\newcommand{\field}{\mathit{fld}}
\newcommand{\method}{\mathit{mtd}}
\newcommand{\str}{\mathit{str}}
\newcommand{\stm}{\mathit{st}}
\newcommand{\prim}{\mathit{prim}}
\newcommand{\lhs}{\mathit{lhs}}
\newcommand{\rhs}{\mathit{rhs}}
\newcommand{\goto}[1]{\texttt{goto}\ #1}
\newcommand{\move}[2]{\texttt{move}\ #1\ #2}
\newcommand{\new}[2]{\texttt{new}\ #1\ #2}
\newcommand{\newarray}[3]{\texttt{newarray}\ #1\ #2\ #3}
\newcommand{\checkcast}[2]{\texttt{checkcast}\ #1\ #2}
\newcommand{\instanceof}[3]{\texttt{instof}\ #1\ #2\ #3}
\newcommand{\invoke}[3]{\texttt{invoke}\ #1\ #2\ #3}
\newcommand{\sinvoke}[3]{\texttt{sinvoke}\ #1\ #2\ #3}
\newcommand{\return}{\texttt{return}}
\newcommand{\ifbr}[3]{\texttt{if}_{\comp}\ #1\ #2\ \texttt{then}\ #3}
\newcommand{\comp}{\varolessthan}
\newcommand{\pc}{\mathit{pc}}
\newcommand{\define}{::=}
\newcommand{\unop}[2]{\texttt{unop}_{\odot}\ #1\ #2}
\newcommand{\arrtype}[1]{\texttt{array}[#1]}
\newcommand{\type}[1]{\texttt{#1}}
\newcommand{\primtype}{\mathit{\tau_{prim}}}
\newcommand{\binop}[3]{\texttt{binop}_{\oplus}\ #1\ #2\ #3}
\newcommand{\subtype}{\leq}
\newcommand{\callstack}{\alpha}
\newcommand{\ind}{\mathit{idx}}
\newcommand{\regval}[1]{\Sigma \llbracket #1 \rrbracket}
\newcommand{\meth}[3]{#1: #2\ \{#3\}}
\newcommand{\true}{\mathtt{true}}
\newcommand{\false}{\mathtt{false}}
\newcommand{\obj}[2]{\{\!| #1; #2 |\!\}}
\newcommand{\arr}[2]{#1[#2]}
\newcommand{\super}{\textit{super}}
\newcommand{\interfaces}{\textit{inter}}
\newcommand{\getst}[1]{\textit{get-stm} (#1)}
\newcommand{\gettype}[2]{\textit{type}_{#1} (#2)}
\newcommand{\lookup}{\textit{lookup}}
\newcommand{\loc}{\mathit{loc}}
\newcommand{\ret}{\mathit{ret}}
\newcommand{\locstate}[4]{\langle #1 \cdot #4 \cdot #2 \cdot #3 \rangle}
\newcommand{\size}{\mathit{len}}
\newcommand{\intent}[2]{\{\!|@ #1; #2 |\!\}}
\newcommand{\defvalue}{\mathbf{0}}
\newcommand{\registers}{\textit{Registers}}
\newcommand{\newintent}[2]{\texttt{newintent}\ #1\ #2}
\newcommand{\result}{\textit{result}}
\newcommand{\locnull}{\texttt{null}}
\newcommand{\void}{\texttt{void}}
\newcommand{\ifeq}[3]{\texttt{if}_{=}\ #1\ #2\ \texttt{then}\ #3}
\newcommand{\sign}{\textit{sign}}
\newcommand{\methsign}[3]{#1 \xrightarrow{#3} #2}
\newcommand{\pointer}[2]{#1_{#2}}

\newcommand{\actconf}[3]{#1 \cdot #2 \cdot #3}
\newcommand{\actstack}{\Omega}
\newcommand{\heap}{H}
\newcommand{\sheap}{S}
\newcommand{\actframe}[4]{\langle #1, #2, #3, #4 \rangle}
\newcommand{\uactframe}[4]{\underline{\actframe{#1}{#2}{#3}{#4}}}
\newcommand{\ocallstack}{\overline{\callstack}}
\newcommand{\actstate}[1]{\textit{#1}}
\newcommand{\finished}{\textit{finished}}
\newcommand{\parent}{\textit{parent}}
\newcommand{\getcb}[2]{\callstack_{#1.#2}}
\newcommand{\lifecycle}{\textit{Lifecycle}}
\newcommand{\actstates}{\textit{ActStates}}
\newcommand{\putextra}[3]{\texttt{put-extra}\ #1\ #2\ #3}
\newcommand{\getextra}[3]{\texttt{get-extra}\ #1\ #2\ #3}
\newcommand{\startact}[1]{\texttt{start-activity}\ #1}
\newcommand{\setresult}{\textit{setResult}}
\newcommand{\finish}{\textit{finish}}
\newcommand{\getintent}{\textit{getIntent}}
\newcommand{\handlers}{\textit{handlers}}
\newcommand{\cb}{\mathit{cb}}
\newcommand{\fintent}{\textit{intent}}
\newcommand{\serval}[1]{\textit{ser}_{\textit{Val}}^{#1}}
\newcommand{\serblock}[1]{\textit{ser}_{\textit{Blk}}^{#1}}

\newcommand{\fact}{\mathsf{f}}
\newcommand{\absual}{\hat{u}}
\newcommand{\absval}{\hat{v}}
\newcommand{\absreg}[3]{\mathsf{R}_{#1}(#2\, ;\, #3)}
\newcommand{\absheap}{\mathsf{H}}
\newcommand{\abssheap}{\mathsf{S}}
\newcommand{\absloc}{\ann}
\newcommand{\apc}{\mathsf{c},\mathsf{m},\mathsf{pc}}
\newcommand{\apcn}{\mathsf{c},\mathsf{m},\mathsf{pc+1}}
\newcommand{\apcp}{\mathsf{c},\mathsf{m},\mathsf{pc'}}
\newcommand{\ainst}[1]{(\!| #1 |\!)_{\pp}}
\newcommand{\acomp}{\hat{\comp}}
\newcommand{\abinop}{\hat{\oplus}}
\newcommand{\aunop}{\hat{\odot}}
\newcommand{\earhs}[1]{\langle\!\langle #1 \rangle\!\rangle_{c,m,\pc}}
\newcommand{\arhs}[1]{\langle\!\langle #1 \rangle\!\rangle_{\pp}}
\newcommand{\prhs}[1]{\mathsf{RHS}_{\spp} (#1)}
\newcommand{\eprhs}[1]{\mathsf{RHS}_{\apc} (#1)}
\newcommand{\absprim}{\widehat{\prim}}
\newcommand{\absobj}[2]{\{\!| #1; #2 |\!\}}
\newcommand{\absarray}[2]{#1[#2]}
\newcommand{\absintent}[2]{\{\!| @#1; #2 |\!\}}
\newcommand{\absblock}{\hat{b}}
\newcommand{\absprog}{\Delta}
\newcommand{\adefvalue}{\hat{\mathbf{0}}}
\newcommand{\absresult}[3]{\mathsf{Res}_{#1}(#2\, ;\, #3)}
\newcommand{\absact}[4]{\mathsf{A}(#1; \fintent : #2, \parent : #3, \result : #4)}
\newcommand{\absgettype}{\widehat{\textit{get-type}}}
\newcommand{\translate}[1]{(\!| #1 |\!)}
\newcommand{\ainstfull}[2]{(\!| #1 |\!)_{#2}}
\newcommand{\rulename}[1]{\textit{#1}}
\newcommand{\absdispatch}{\mathsf{I}}
\newcommand{\abslookup}{\widehat{\textit{lookup}}}
\newcommand{\const}[1]{\mathsf{#1}}

\newcommand{\taint}{\mathit{h}}
\newcommand{\public}{\mathsf{public}}
\newcommand{\secret}{\mathsf{secret}}
\newcommand{\taintf}{\hslash}
\newcommand{\ataintf}{\hat{\hslash}}

\newcommand{\rfprim}{\beta_{\textit{Prim}}}
\newcommand{\rfloc}{\beta_{\textit{Loc}}}
\newcommand{\rfval}{\beta_{\textit{Val}}}
\newcommand{\rfblock}{\beta_{\textit{Blk}}}
\newcommand{\rfcall}{\beta_{\textit{Call}}}
\newcommand{\rfheap}{\beta_{\textit{Heap}}}
\newcommand{\rfstat}{\beta_{\textit{Stat}}}
\newcommand{\rflconf}{\beta_{\textit{Lcnf}}}
\newcommand{\rfconf}{\beta_{\textit{Cnf}}}
\newcommand{\rfastk}{\beta_{\textit{Stk}}}
\newcommand{\rflocstate}{\beta_{\textit{Lst}}}
\newcommand{\rfframe}{\beta_{\textit{Frm}}}
\newcommand{\rfdispatch}[1]{\beta_{\textit{Pact}}^{#1}}
\newcommand{\poprim}{\sqsubseteq_{\textit{Prim}}}
\newcommand{\poval}{\sqsubseteq_{\textit{Val}}}
\newcommand{\poblk}{\sqsubseteq_{\textit{Blk}}}
\newcommand{\poseq}{\sqsubseteq_{\textit{Seq}}}
\newcommand{\sinks}{\textit{Sinks}}
\newcommand{\sources}{\textit{Sources}}
\newcommand{\methconf}[5]{#5 \cdot #1 \cdot #2 \cdot #3 \cdot #4}
\newcommand{\ann}{\lambda}
\newcommand{\astart}[1]{in(#1)}
\newcommand{\pp}{\mathit{pp}}
\newcommand{\spp}{\mathsf{pp}}
\newcommand{\bvsize}[1]{\textit{size}^{#1}}
\newcommand{\serialized}{\Gamma}
\newcommand{\newpointer}[1]{\nu(#1)}

\newcommand{\slocstate}[3]{\langle #1 \cdot #2 \cdot #3 \rangle}
\newcommand{\smethconf}[4]{#1 \cdot #2 \cdot #3 \cdot #4}


\maketitle

\begin{abstract}
  We present \tool, a new tool for the static analysis of information
  flow properties in Android applications. The core idea underlying
  \tool is to use Horn clauses for soundly abstracting the semantics of
  Android applications and to express security properties as a set of
  proof obligations that are automatically discharged by an
  off-the-shelf SMT solver. This approach makes it possible to fine-tune the
  analysis in order to achieve a high degree of precision while still using off-the-shelf verification tools, thereby 
  leveraging the recent advances  in this field. As a
  matter of fact,   \tool outperforms
  state-of-the-art Android static analysis tools on benchmarks
  proposed by the community. Moreover, \tool is the first static analysis
  tool for Android to come with a formal proof of soundness, which
  covers the core  of the analysis technique: besides yielding correctness assurances,
  this proof allowed us to identify some critical corner-cases that affect 
  the soundness guarantees provided by some of the previous static analysis 
  tools for Android.
\end{abstract}

\section{Introduction}
\label{sec:intro}
The Android platform is by far the most popular choice for mobile
devices nowadays, with billions of applications routinely installed on
a massive number of different phones and tablets. Given this
increasing popularity, personal information and other
sensitive data stored on Android devices constitute an attractive
target for breaching users' privacy at scale by malicious application
developers. Information flow control frameworks for Android have thus
emerged as a prominent research direction, with several different
proposals spanning from dynamic
analysis~\cite{EnckGHTCCJMS14,JiaAFBSFKM13,TrippR14,HornyackHJSW11} to static
analysis~\cite{ZhaoO12,YangY12,MannS12,GiblerCEC12,Kim12,LuLWLJ12,ArztRFBBKTOM14,GordonKPGNR15,LiEtAl15}. Static analysis is particularly appealing for information flow control,
given its ability to provide full coverage of all the possible execution paths
and the possibility to be employed in the vetting phase, i.e., before
the  application is uploaded onto the Google Play store.  

The most recent works in this
area~\cite{ArztRFBBKTOM14,GordonKPGNR15,LiEtAl15,WeiROR14} are
impressive in their efforts to support a significant fragment of the Android
platform. Most of them leverage existing static analysers by encoding
Android applications in a suitable format, e.g.,  FlowDroid~\cite{ArztRFBBKTOM14},  
DroidSafe~\cite{GordonKPGNR15}, and IccTA~\cite{LiEtAl15} use Soot~\cite{Vallee-Rai:2000},
while CHEX~\cite{LuLWLJ12} uses Wala~\cite{Wala}. Observing that
existing static analysers come with intrinsic limitations that limit
the precision of the analysis (e.g.,
Soot and Wala do not  calculate all objects' points-to information in a
both flow- and context-sensitive way), Amandroid~\cite{WeiROR14}
relies on a dedicated data-flow analysis algorithm. 

Despite all this progress and sophisticated machinery, none of these tools
achieves a satisfactory degree of soundness: even on benchmarks
written by the community and consisting of simple programs (i.e.,
Droidbench~\cite{ArztRFBBKTOM14}), for which
the ground truth is known, all
existing tools miss several malicious leaks (false negatives). This, along with
the fact that none of these tools
comes with a formal model or soundness proof, makes one
wonder how accurately these analyses capture all the
subtleties of the Android execution model, which is far from being
trivial~\cite{PayetS14}, and to which extent their results are reliable
on real-life applications, for which the ground truth is not known. 

Furthermore, the lack of precise and fully
documented analysis definitions complicates the comparison between
different approaches: for instance, there is no universal agreement on
a single notion of object-sensitivity~\cite{SmaragdakisBL11}, though
object-sensitivity has been recognized as crucial to support a precise
analysis of real-world Android applications~\cite{ArztRFBBKTOM14}. Hence, at
the time of writing, the only way to grasp the relative strengths and
weaknesses of different static analysis tools for Android applications
relies on an hands-on testing on some common benchmark and a source 
code inspection of their implementation.

\medskip
\noindent
\textbf{Our contributions.} 
We present a fresh approach to the static   analysis of
Android applications, i.e., a data-flow analysis based on \emph{Horn clause
  resolution}~\cite{BjornerMR12}. The core idea is to soundly abstract the semantics of
Android applications into a set of Horn clauses and to formulate security
properties as a set of proof obligations, which can be automatically discharged
by off-the-shelf SMT solvers. In particular:

\begin{itemize}
\item We prove the soundness of our analysis against a rigorous formal model
  of a large fragment of the Android ecosystem, covering Dalvik bytecode,
  the event-driven nature of the activity lifecycle, and inter-component 
  communication. While elaborating the proof, we identified a few critical 
  corner-cases that affect the soundness guarantees provided by some of 
  previous static analysis tools for Android. We believe that this formal model
  may constitute a foundational framework, serving as a starting and comparison
  point for future work in the field; 

\item We fine-tune the Horn clause generation in order to 
  optimize precision and efficiency, while retaining
  soundness. Being a data-flow analysis rather than a pure taint analysis,
  our solution statically approximates run-time values, in contrast to
  most of the previous works in the field~\cite{ArztRFBBKTOM14,GordonKPGNR15,LiEtAl15}. This boosts
  the precision of the analysis: for instance, it makes it possible to
  statically determine whether a conditional branch will never be
  taken at runtime and ignore it. A salient feature of our approach is
  the usage of SMT solving to discharge proof obligations. From an
  engineering point of view, this allows one to fine-tune the analysis while still building on
  off-the-shelf verification tools, thereby leveraging the
  continuous advances in this field. 

\item We develop a tool, called \tool, which implements the
  analysis described in the formal model and complements it in order
  to support additional Android features, such as reflection,
  exceptions, and threading. \tool automatically generates Horn
  clauses from the application bytecode and relies on
  the state-of-the-art SMT solver Z3~\cite{MouraB08} for discharging
  proof obligations\footnote{It would  be possible to discharge proof
    obligations by using at the same time different SMT
    solvers, since each of them might perform best on a certain class
    of queries. We did not find it necessary in our current
    experiments, but we plan to implement this feature in the
    future.}. 

\item  We conduct a performance evaluation
  on Droidbench, a collection of 120 programs written by the
  community, comparing \tool with IccTA~\cite{LiEtAl15} (an
  extension of FlowDroid~\cite{ArztRFBBKTOM14} to inter-component communication), Amandroid~\cite{WeiROR14}
and DroidSafe~\cite{GordonKPGNR15}.   \tool outperforms the
competitors in terms of sensitivity (i.e., soundness) and performance,
while retaining a high 
specificity (i.e., precision): 
\tool is the only tool that identifies all the explicit information flows, it
  exhibits just one more false positive than Amandroid (the most
  accurate tool), and it is one order of magnitude faster than IccTA and
 AmanDroid,  and two orders of magnitude faster than DroidSafe. Furthermore, we show 
that \tool{} scales well to real-life applications from Google Play by a 
 comparative evaluation  on the two largest applications from the Google Play Top 30, 
 i.e., Candy Crash Soda Saga and Facebook, which pose significant problems to existing tools.
\end{itemize} 

\iffull \else The tool as well as an extended version of this paper with a complete formalisation and proofs are available online~\cite{full}. \fi


\section{Design and Motivations}
\label{sec:design}
Static information flow control for Android applications is a mature research area nowadays~\cite{ZhaoO12,YangY12,MannS12,GiblerCEC12,Kim12,LuLWLJ12}, with IccTA~\cite{LiEtAl15} (an extension of  FlowDroid~\cite{ArztRFBBKTOM14} to inter-component communication), AmanDroid~\cite{WeiROR14} and DroidSafe~\cite{GordonKPGNR15} representing the state-of-the-art in this field. Although all these proposals are impressive projects, which significantly advanced the area of information flow control for Android applications, they all have important limitations, motivating the need for novel research proposals. 

We make this need apparent by focussing on two important design choices where these tools differ: \emph{value-sensitivity} and \emph{flow-sensitivity}. It is instructive to highlight the import of these choices in terms of both the soundness and the precision of the resulting static analysis. Table~\ref{tab:design} summarizes the design choices of the tools we consider, including ours.

\begin{table}[htb]
\begin{center}
\begin{tabular}{c|c|c|c|c|}
\cline{2-5}
& \emph{IccTA} & \emph{AD} & \emph{DS} & \emph{HD} \\
\cline{1-5}
\multicolumn{1}{|c|}{\emph{Value-sensitivity}} & no & yes & no & yes \\ 
\cline{1-5}
\multicolumn{1}{|c|}{\emph{Flow-sensitivity}} & yes & yes & no & partial \\
\cline{1-5}
\end{tabular}
\end{center}
\caption{Design Choices for Static Analysis Tools}
\label{tab:design}
\end{table}

\subsection{Value-sensitivity}
Value-sensitivity is the ability of a static analysis to approximate runtime values and use this information to improve precision, e.g., by skipping unreachable program branches~\cite{NielsonNH99}. Concretely, consider the following code:
\begin{verbatim}
int x = 0;
for (int y = 0; y <= 10; y++) { x++; }
TelephonyManager tm = ...
String imei = tm.getDeviceId();
if (x == 0) { leak(imei); }
\end{verbatim}
Though this code is perfectly safe, all the existing tools (IccTA, AmanDroid and DroidSafe) will identify it as leaky. IccTA and DroidSafe conservatively assume all the program points to be potentially reachable. Even AmanDroid raises a false alarm for this code, though it internally implements a dedicated data-flow analysis~\cite{WeiROR14}. 

Besides this simple example, there are many reasons why real-world static analysis tools for Android applications should be value-sensitive to be practically useful. First, several features of Java and the Android APIs, most notably \emph{reflection} and \emph{dictionary-like} containers, e.g., intents and bundles, need value-sensitivity to be analysed precisely. Second, the loss of precision entailed by value-insensitivity may creep and interact badly with other desirable features of the static analysis, e.g., \emph{context-sensitivity}, which has been deemed as crucial by previous studies~\cite{ArztRFBBKTOM14,GordonKPGNR15}. 

Context-sensitivity is the ability of the analysis to compute
different static approximations upon different method calls. To
understand why the benefits of context-sensitivity can be voided by
value-insensitivity, consider the following method, where we assume to
know a valid upper bound for the GPS location values:

\begin{verbatim}
void m (double x, double y) {
  if (x <= MAX_X && y <= MAX_Y)
    ...
  else
    leak("Invalid location:" + x + y);
}
\end{verbatim} 
Context-insensitive static analyses would detect a dangerous information flow whenever the method \texttt{m} is invoked at two different program points and one of these invocations provides the location of the device in the actual parameters, while the other one provides an invalid location. The reason is that the method \texttt{m} would be analysed only once, hence the static analysis would detect that both public and confidential values may reach a sink. Conversely, a context-sensitive analysis potentially has the ability to discriminate between the two methods invocations and be precise, but the lack of value-sensitivity would necessarily lead to the detection of a non-existent information flow.

Finally, it is worth noticing that  value-sensitivity is crucial to support 
security-relevant, value-dependent security queries (e.g., ``Is the credit card 
number sent on HTTP rather than on HTTPS?'' or ``Is the picture actually 
uploaded on Facebook, as opposed to some other untrusted website?'').

\subsection{Flow-sensitivity}
Flow-sensitivity is the ability of a static analysis to take the order of statements into account and compute different approximations at different program points~\cite{NielsonNH99}. To understand its importance, consider the following code:
\begin{verbatim}
TelephonyManager tm = ...
String imei = tm.getDeviceId();
imei = new String("empty");
leak(imei);
\end{verbatim}
Though the code above is safe, the flow-insensitive analysis implemented in DroidSafe will identify it as leaky, since the variable \texttt{imei} does contain a secret information at some program point. Conversely, both FlowDroid and AmanDroid will correctly deem the program as safe.

Clearly, it is tempting to target a flow-sensitive information flow analysis tool to achieve a higher level of precision, but, as pointed out by the authors of DroidSafe~\cite{GordonKPGNR15}, flow-sensitivity is very hard to get right for Android applications, due to their massive use of asynchronous callbacks. Both FlowDroid and AmanDroid suggest to tackle this problem by introducing a \emph{dummy main method} emulating each possible interleaving of the callbacks defining the application life-cycle. Unfortunately, it is difficult to ensure that the dummy main method construction is accurate and comprehensive, which leads to missing malicious information flows~\cite{GordonKPGNR15}. 

\subsection{HornDroid}
Our tool, \tool, targets a \emph{sound} and \emph{practical} information flow analysis for Android applications. We report on the design choices we made to hit the sweet spot between these two potentially conflicting requirements.

\tool{} implements a \emph{value-sensitive} information flow analysis. As anticipated, value-sensitivity is crucial to support a practically useful analysis of real-world applications. The analysis implemented in \tool{} is reminiscent of \emph{abstract interpretation}, whereby the operational semantics of a program is over-approximated by a computable abstract semantics. As it is customary for abstract interpretation, the design of the analysis is parametric with respect to the choice of a set of \emph{abstract domains}, defining how runtime values are statically approximated: one can then fine-tune the precision of the analysis by testing different abstract domains. To ensure the scalability of our value-sensitive analysis, the abstract semantics implemented in \tool{} is based on \emph{Horn clauses}, whose efficient resolution is supported by state-of-the-art SMT solvers~\cite{BjornerMR12}.

\tool{} performs a \emph{flow-sensitive} information flow analysis on the registers employed by the Dalvik Virtual Machine, while implementing a \emph{flow-insensitive} analysis for callback methods and heap locations. This is crucial to preserve the precision of the analysis, without sacrificing soundness. We already mentioned that previous studies highlighted that flow-sensitive analyses may easily produce unsound results, due to the challenges of predicting all the possible orderings of the Android callbacks~\cite{GordonKPGNR15}. Moreover, while carrying out the soundness proof for \tool, we realized that \emph{static fields} are particularly delicate to treat in a flow-sensitive fashion. The reason is that static fields provide a way to implement a shared memory between otherwise memory-isolated components running in the same application. Given that the execution order of different Android components is extremely hard to predict, due to their callback-driven nature, it turns out that flow-insensitivity for static fields is in practice needed for soundness. Indeed, since static fields can be used to exchange pointers to heap locations, a sound flow-sensitive analysis for heap locations is in general hard to achieve. Our soundness proof, instead, confirms that flow-sensitivity can be implemented for the registers employed by the Dalvik Virtual Machine without missing any malicious information flow.


\section{Operational Semantics}
\label{sec:activity}
We base our technical development on \sem, a formal model of the Android semantics obtained by extending the $\mu$-Dalvik calculus~\cite{JeonMF12} with a complete characterisation of the activity-specific aspects of the Android platform~\cite{PayetS14}.

\subsection{Background and Scope}
Android applications are developed in Java and then compiled to a custom bytecode format called \emph{Dalvik}, which is run by the Dalvik Virtual Machine (DVM). Unlike Java VMs, which are stack machines, the DVM adopts a register-based architecture.
Android applications are different from standard Java programs, since
they are structured in \emph{components} of four different types:
activities, services, content providers and broadcast
receivers~\cite{Android}. These components represent distinct entry
points of the Android framework into the application. Hence, the
operational behaviour of an Android application does not simply amount
to the sequential execution of its bytecode implementation, but it
heavily relies on callbacks from the Android framework, as a reaction
to user inputs, system events, or inter-component
communication. Different Android components, either in the same
application or from different applications, can communicate by
exchanging \emph{intents}, i.e., dictionary-like messaging
objects. Intents may be sent either to a specific component
(\emph{explicit} intents) or to any component which declares the will
of providing a given functionality (\emph{implicit} intents).  

In our formal model we consider Android applications consisting of
activities only. We focus on activities, since a tested semantics is
available for them and because they exhibit the most complicated
life-cycle among all the component types~\cite{PayetS14}. Also, we
only model intra-application communication based on explicit
intents: implicit intents are mostly, if not only, used for inter-application
messages. As we discuss in Section~\ref{sec:experiments}, \sem does not
cover all the Android features supported by \tool{}: the purpose of \sem is
ensuring that the design principles at the core of \tool{} are sound and
that most of the Android-specific subtleties have been taken into due account. 

\subsection{Syntax}
We write $(r_i)^{i \leq n}$ for the sequence $r_1,\ldots,r_n$. If the length of the sequence is immaterial, we just write $r^*$ and we still let $r_j$ stand for its $j$-th element. We represent the empty sequence with a dot ($\cdot$). We let $r^*[j \mapsto r']$ be the sequence obtained from $r^*$ by replacing its $j$-th element with $r'$. A \emph{partial map} is a sequence of key-value bindings $(k_i \mapsto v_i)^*$, where all the keys $k_i$ are pairwise distinct. Given a partial map $M$, let $\dom(M)$ stand for the set of its keys and let $M(k) = v$ whenever the binding $k \mapsto v$ occurs in $M$. We identify partial maps which are identical up to the order of their key-value bindings.

Table~\ref{tab:dalvik} provides the syntax of \sem{} programs. It is an extension of the original $\mu$-Dalvik syntax~\cite{JeonMF12} with a few additional statements modelling method calls to Android APIs used for inter-component communication.

\begin{table}[t]
\[\begin{array}{lll}
P & \define & \class^*  \\
\class & \define & \cls{c}{c'}{c^*}{\field^*}{\method^*}  \\
\primtype & \define & \type{bool} ~|~ \type{int} ~|~ \dots \\
\tau & \define & c ~|~ \primtype ~|~ \arrtype{\tau} \\ 
\field & \define & f: \tau \\
\method & \define & \meth{m}{\methsign{\tau^*}{\tau}{n}}{\stm^*} \\\\
\stm & \define & \goto{\pc} \\
 & | & \move{\lhs}{\rhs} \\
 & | & \ifbr{r_1}{r_2}{\pc} \\
 & | & \unop{r_d}{r_s} \\
 & | & \binop{r_d}{r_1}{r_2} \\
 & | & \new{r_d}{c} \\
 & | & \newarray{r_d}{r_l}{\tau} \\
 & | & \checkcast{r_s}{\tau} \\
 & | & \instanceof{r_d}{r_s}{\tau} \\
 & | & \invoke{r_o}{m}{r^*} \\
 & | & \sinvoke{c}{m}{r^*} \\
 & | & \return \\
 & | & \newintent{r_i}{c} \\
 & | & \putextra{r_i}{r_k}{r_v} \\
 & | & \getextra{r_i}{r_k}{\tau} \\
 & | & \startact{r_i}
\\\\
r & \in & \registers \\
\pc & \in & \mathbb{N} \\
\oplus & \define & + ~|~ - ~|~ \dots \\
\odot & \define & - ~|~ \neg ~|~ \dots \\
\comp & \define & < ~|~ > ~|~ \dots \\
\prim & \define & \true ~|~ \false ~|~ \dots \\
\lhs & \define & r \\
& | & r[r] \\
& | & r.f \\
& | & c.f \\
\rhs & \define & \lhs \\
& | & \prim
\end{array}
\]
\caption{\sem{} Syntax}
\label{tab:dalvik}
\end{table}

A \sem{} program $P$ is a sequence of classes $\class^*$, which in turn are defined by a class name $c$, a direct super-class $c'$, some implemented interfaces $c^*$, and a number of fields $\field^*$ and methods $\method^*$. Field declarations $f: \tau$ include the field name $f$ and its type $\tau$, while method declarations $\meth{m}{\methsign{\tau^*}{\tau}{n}}{\stm^*}$ include the method name $m$, the argument types $\tau^*$, the return type $\tau$, and the method body $\stm^*$. The annotation $n$ on top of the arrow tracks the number of local registers used by the method, which is statically known in Dalvik.

We briefly discuss below the statements of the language. An
unconditional branch $\goto{\pc}$ sets the program counter to
$\pc$. The statement $\move{\lhs}{\rhs}$ moves the right-hand side
$\rhs$ into the left-hand side $\lhs$: here, $\lhs$ may be a register
$r$, an array cell $r_1[r_2]$, an object field $r.f$, or a static field
$c.f$; $\rhs$ may be any of these elements or a constant. A
conditional branch $\ifbr{r_1}{r_2}{\pc}$ compares the content of two
registers $r_1$ and $r_2$ using the comparison operator $\comp$ and
sets the program counter to $\pc$ if the check is successful,
otherwise it moves to the next instruction. We then have unary and
binary operations, represented by $\unop{r_d}{r_s}$ and
$\binop{r_d}{r_1}{r_2}$ respectively, where $r_d$ is the destination
register where the result of the operation must be stored and the
other registers contain the operands. Object creation is modelled by
$\new{r_d}{c}$, which creates an object of class $c$ and stores a
pointer to it in $r_d$; array creation is similarly handled by
$\newarray{r_d}{r_l}{\tau}$, where $r_d$ is the destination register
where the pointer to the new array must be stored, $r_l$ contains the
array length and $\tau$ specifies the type of the array cells. The
type cast statement $\checkcast{r_s}{\tau}$ checks whether the
register $r_s$ contains a pointer to an object of type $\tau$ and it
moves to the next instruction if this is the case, otherwise it stops
the execution\footnote{The corresponding Dalvik opcodes would raise an
exception, but we do not model exceptions in our formalism.}. The
statement $\instanceof{r_d}{r_s}{\tau}$ stores $\true$ in $r_d$ if
$r_s$ points to an object of type $\tau$, otherwise it stores
$\false$. A method invocation $\invoke{r_o}{m}{r^*}$ calls the method
$m$ on the receiver object pointed by $r_o$, passing the values in the
registers $r^*$ as actual arguments. The invocation of static methods
is modelled by $\sinvoke{c}{m}{r^*}$. The $\return$ statement has no
argument, rather there is a special register $r_{\ret}$ for holding
return values: the return value must be moved to $r_{\ret}$ by the
callee before calling $\return$.  

The last four statements are used to model inter-component
communication. Intent creation is modelled by $\newintent{r_i}{c}$,
which creates an intent for the activity $c$ and stores a pointer to
it in $r_i$. The statement $\putextra{r_i}{r_k}{r_v}$ adds to the
intent pointed by $r_i$ a new key-value binding $k \mapsto v$, where
$k$ and $v$ are the contents of $r_k$ and $r_v$ respectively. The
statement $\getextra{r_i}{r_k}{\tau}$ retrieves from the intent
pointed by $r_i$ the value bound to key $k$, where $k$ is the content
of $r_k$, provided that this value has type $\tau$. Finally,
$\startact{r_i}$ sends the intent pointed by $r_i$, thus starting a
new activity. Throughout the paper, we only consider \emph{well-formed} programs.

\begin{definition}
A program $P$ is \emph{well-formed} iff: (1) all its class names are pairwise distinct, (2) for each of its classes, all the field names are pairwise distinct, and (3) for each of its classes, all the method names are pairwise distinct.
\end{definition}

Notice that the last condition of the definition above is not restrictive, since overloading resolution is performed at compile time in Java~\cite{JavaSpec} and Dalvik bytecode thus identifies methods through their signature, rather than their name. In our formalism, we then suppose that method names are tagged with some distinctive information drawn from their signature, so that we can identify each method of a given class just by its name. Notice that two different classes can still define two methods with the same name, which is important to model dynamic dispatching.

From now on, we focus our attention on some well-formed program $P = \class^*$. Most of the definitions we present in the paper depend on $P$, but we do not make this dependence explicit in the notation to keep it lighter.

\subsection{Dalvik Semantics}
\label{sec:dalvik}
Table~\ref{tab:dalvik-domains} defines the semantic domains employed by the operational semantics of \sem. Values include primitive values and \emph{locations}, i.e., pointers to heap elements extended with an annotation $\ann$. Annotations have no semantic import and are only needed for our static analysis: we will discuss their role in Section~\ref{sec:statics}. 

\begin{table*}[htb]
\begin{mathpar}
\begin{array}{llcl}
\text{Pointers} & p & \in & \textit{Pointers} \\
\text{Program points} & \pp & \define & c,m,\pc \\
\text{Annotations} & \ann & \define & \pp ~|~ c ~|~ \astart{c} \\
\text{Locations} & \ell & \define & \pointer{p}{\ann} \\
\text{Values} & u,v & \define & \prim ~|~ \ell \\
\text{Registers} & R & \define & (r \mapsto v)^* \\
\text{Local states} & L & \define & \slocstate{\pp}{\stm^*}{R} \\
\text{Call stacks} & \callstack & \define & \varepsilon ~|~ L :: \callstack \\
\text{Pending activity stacks} & \pi & \define & \varepsilon ~|~ i :: \pi \\
\end{array}
\begin{array}{llll}
\text{Objects} & o & \define & \obj{c}{(f_{\tau} \mapsto v)^*} \\
\text{Arrays} & a & \define & \arr{\tau}{v^*} \\
\text{Intents} & i & \define & \intent{c}{(k \mapsto v)^*} \\
\text{Memory blocks} & b & \define & o ~|~ a ~|~ i \\
\text{Heaps} & \heap & \define & (\ell \mapsto b)^* \\
\text{Static heaps} & \sheap & \define & (c.f \mapsto v)^* \\
\text{Local configurations} & \Sigma & \define & \smethconf{\callstack}{\pi}{\heap}{\sheap}
\end{array}
\end{mathpar}
\caption{\sem{} Semantic Domains}
\label{tab:dalvik-domains}
\end{table*}

A \emph{local configuration} $\Sigma =
\smethconf{\callstack}{\pi}{\heap}{\sheap}$ represents the state of a
specific activity. It includes a call stack $\callstack$, a pending
activity stack $\pi$, a heap $\heap$, and a static heap $\sheap$. A
call stack $\callstack$ is a list of \emph{local states}, which is
populated upon method invocation. Each local state includes: (1) a
program point $\pp = c,m,\pc$, where $c$ and $m$ identify the invoked
method, while $\pc$ points to the next instruction to execute; (2) a list
of statements $\stm^*$, modelling the method body; and (3) a map $R$ 
binding local registers to their current value. 

A pending activity stack $\pi$ is a list of intents, which are treated
as (untyped) dictionaries in our formalism. As anticipated, for the
sake of simplicity, we only consider
explicit intents in the formalization, i.e., intents which are meant to be delivered
to an activity of a given class $c$: this class is specified after the
\lq at\rq\ symbol (@) in the intent syntax\footnote{Extending the formalism
to include implicit intents would not be difficult, but this would introduce
non-determinism on the choice of the receiving activity, thus making
the presentation harder to follow.}. We use $\pi$ to keep track of
which activities have been started by the activity modelled by the
local configuration.

Finally, a heap $\heap$ is a mapping between locations and memory
blocks, where each block is either an object, an array or an
intent. Object fields are annotated with their static type, though we
typically omit this annotation when it is unimportant. The static heap
$\sheap$ simply binds static fields to their corresponding value. 

The small-step operational semantics of \sem{} is defined by a reduction relation $\Sigma \rightsquigarrow \Sigma'$. Reduction takes place by fetching the next statement to execute, based on the program counter of the top-most local state of the call stack in $\Sigma$, and by running it to produce $\Sigma'$. The definition of the reduction relation is lengthy, but unsurprising, and it is given in \iffull Appendix~\ref{sec:bytecode}. \else the full version~\cite{full}. \fi The only point worth noticing here is that, when a new memory block is created, e.g., by \texttt{new}, the corresponding pointer to the heap is annotated with the program point $c,m,\pc$ where creation takes place.

\subsection{Activity Semantics}

The operational behaviour of an activity does not depend only on its bytecode implementation, but also on external events, like user inputs and system callbacks. The event-driven nature of Android applications gives rise to highly non-deterministic executions, which are not trivial to approximate correctly by static analysis.

\subsubsection{Formalizing Activities}

We start by introducing a formal notion of activity.

\begin{definition}
A class $\class$ is an \emph{activity class} if and only if $\class = \cls{c}{c'}{c^*}{\field^*}{\method^*}$ for some $c' \subtype \texttt{Activity}$. An \emph{activity} is an instance of an activity class. We stipulate that each activity has the following fields: (1) $\finished$: a boolean flag stating whether the activity has finished or not; (2) $\fintent$: a pointer to the intent which started the activity; (3) $\result$: a pointer to an intent storing the result of the activity computation; and (4) $\parent$: a pointer to the parent activity, i.e., the activity which started the present one.
\end{definition}

We require that each activity has a (possibly empty) set of \emph{event handlers} for user inputs: given an activity class $c$, we let $\handlers(c) = \{m_1,\ldots,m_n\}$ be the set of the names of the methods of $c$ which may be dispatched when some user input event occurs. We assume a set of activity states $\actstates$ and a relation $\lifecycle \subseteq \actstates \times \actstates$ defining the state transitions admitted by the activity lifecycle~\cite{PayetS14}. We assume that each activity class $c$ has a set of callbacks for each activity state $s$, whose names are returned by a function $\cb(c,s)$; for the $\actstate{running}$ state we let $\cb(c,\actstate{running}) = \handlers(c)$, i.e., when an activity is running, any callback set for user inputs may be dispatched.

We then extend the syntax of \sem{} with the elements in Table~\ref{tab:ext-dalvik}. A \emph{frame} $\varphi$ includes a location $\ell$ pointing to an activity, a corresponding activity state $s$, a pending activity stack $\pi$ and a call stack $\callstack$. Frames are organized in an \emph{activity stack} $\actstack$, modelling different activities executing in the same application: a single frame in $\actstack$ has priority of execution and is underlined. A \emph{configuration} $\Psi$ includes an activity stack $\actstack$, a heap $\heap$ and a static heap $\sheap$.

\begin{table}[ht]
\[
\begin{array}{llcl}
\text{Activity states} & s & \in & \actstates \\
\text{Frames} & \varphi & \define & \actframe{\ell}{s}{\pi}{\callstack} ~|~ \uactframe{\ell}{s}{\pi}{\callstack} \\
\text{Activity stacks} & \actstack & \define & \varphi ~|~ \varphi :: \actstack \\
\text{Configurations} & \Psi & \define & \actconf{\actstack}{\heap}{\sheap}
\end{array}
\]
\textbf{Convention:} each activity stack $\actstack$ contains at most one active (underlined) frame.
\caption{Extensions to the Syntax of \sem}
\label{tab:ext-dalvik}
\end{table}

\subsubsection{Reduction Rules}
Before presenting the formal semantics, we need to introduce some additional definitions. We start with the notion of \emph{callback stack}, identifying the admissible format of a call stack for new frames pushed on the activity stack upon the invocation of a callback from the Android system. Let $\sign(c,m) = \methsign{\tau^*}{\tau}{n}$ iff there exists a class $\class_i$ such that $\class_i = \cls{c}{c'}{c^*}{\field^*}{\method^*, \meth{m}{\methsign{\tau^*}{\tau}{n}}{\stm^*}}$. Let then $\lookup$ stand for a \emph{method lookup} function such that $\lookup(c,m) = (c',\stm^*)$ iff: (1) $c'$ is the class defining the method which is dispatched when $m$ is invoked on an object of type $c$, and (2) $\stm^*$ is the method body. 

\begin{definition}
Given a location $\ell$ pointing to an activity of class $c$, we let $\getcb{\ell}{s}$ stand for an arbitrary \emph{callback stack} for state $s$, i.e., any call stack $\slocstate{c',m,0}{\stm^*}{R} :: \varepsilon$, where $(c',\stm^*) = \lookup(c,m)$ for some $m \in \cb(c,s)$, $\sign(c',m) = \methsign{\tau_1,\ldots,\tau_n}{\tau}{\loc}$ and:
\[ 
R = ((r_i \mapsto \defvalue)^{i \leq \loc}, r_{\loc+1} \mapsto \ell, (r_{\loc+1+j} \mapsto v_j)^{j \leq n}),
\]
for some values $v_1,\ldots,v_n$ of the correct type $\tau_1,\ldots,\tau_n$.
\end{definition}

In the definition, we let $\defvalue$ be the default value for local registers. There is just one default value for registers in the model, since registers are untyped in Dalvik. In the following, it is also convenient to presuppose for each type $\tau$ the existence of a a default value $\defvalue_{\tau}$, used to initialize fields of type $\tau$ upon object creation.

A tricky aspect of the operational semantics of activities, which has never been formalized before, is the \emph{serialization} of objects upon inter-component communication. Different activities may exchange objects using intents, but these objects are never passed by reference: rather, they are serialized at the sender side and a copy of them is created at the receiver side. The intent itself is serialized upon communication. We formalize this serialization routine by two mutually recursive functions $\serval{\heap}(v) = (v',\heap')$ and $\serblock{\heap}(b) = (b',\heap')$, returning a serialized copy of their argument and a new heap where all the pointers created in the serialization process have been instantiated correctly. We refer to Table~\ref{tab:activity-semantics} below for the definition of the two functions. Their definition uses a set of pointers $\Gamma$ to keep track of which pointers have already been followed in the serialization process, so as to allow the serialization of memory blocks including self-references.

Finally, the operational semantics requires the next definition of \emph{successful} call stack. A successful call stack is the call stack of an activity which has completed its computation.

\begin{definition}
A call stack $\callstack$ is \emph{successful} if and only if $\callstack = \slocstate{\pp}{\return}{R} :: \varepsilon$ for some $\pp$ and $R$. We let $\ocallstack$ range over successful call stacks.
\end{definition}

Now we have all the ingredients to define the formal semantics of
activities, which is given by the reduction rules in
Table~\ref{tab:activity-semantics}. As anticipated, the rules closely
follow previous work by Payet and Spoto~\cite{PayetS14}, which we
extend to provide a more accurate account of inter-component
communication by modelling value-passing based on a serialization routine. 
We give a short explanation of all the rules, we refer
to~\cite{PayetS14} for a longer description. 

\begin{table*}[p]\small
\begin{mathpar}
\inferrule[(A-Active)]
{\smethconf{\callstack}{\pi}{\heap}{\sheap} \rightsquigarrow \smethconf{\callstack'}{\pi'}{\heap'}{\sheap'}}
{\actconf{\actstack :: \uactframe{\ell}{s}{\pi}{\callstack} :: \actstack'}{\heap}{\sheap} \Rightarrow \actconf{\actstack :: \uactframe{\ell}{s}{\pi'}{\callstack'} :: \actstack'}{\heap'}{\sheap'}}

\inferrule[(A-Deactivate)]
{}
{\actconf{\actstack :: \uactframe{\ell}{s}{\pi}{\ocallstack} :: \actstack'}{\heap}{\sheap} \Rightarrow \actconf{\actstack :: \actframe{\ell}{s}{\pi}{\ocallstack} :: \actstack'}{\heap}{\sheap}}

\inferrule[(A-Step)]
{(s,s') \in \lifecycle \\
\pi \neq \varepsilon \Rightarrow (s,s') = (\actstate{running},\actstate{onPause}) \\
\heap(\ell).\finished = \true \Rightarrow (s,s') \in \{(\actstate{running},\actstate{onPause}),(\actstate{onPause},\actstate{onStop}),(\actstate{onStop},\actstate{onDestroy})\}}
{\actconf{\actframe{\ell}{s}{\pi}{\ocallstack} :: \actstack}{\heap}{\sheap} \Rightarrow \actconf{\uactframe{\ell}{s'}{\pi}{\getcb{\ell}{s'}} :: \actstack}{\heap}{\sheap}}

\inferrule[(A-Destroy)]
{\heap(\ell).\finished = \true}
{\actconf{\actstack :: \actframe{\ell}{\actstate{onDestroy}}{\pi}{\ocallstack} :: \actstack'}{\heap}{\sheap} \Rightarrow \actconf{\actstack :: \actstack'}{\heap}{\sheap}}

\inferrule[(A-Back)]
{\heap' = \heap[\ell \mapsto \heap(\ell)[\finished \mapsto \true]]}
{\actconf{\actframe{\ell}{\actstate{running}}{\varepsilon}{\ocallstack} :: \actstack}{\heap}{\sheap} \Rightarrow \actconf{\actframe{\ell}{\actstate{running}}{\varepsilon}{\ocallstack} :: \actstack}{\heap'}{\sheap}}

\inferrule[(A-Replace)]
{\heap(\ell) = \obj{c}{(f_{\tau} \mapsto v)^*,\finished \mapsto u} \\ 
o = \obj{c}{(f_{\tau} \mapsto \defvalue_{\tau})^*,\finished \mapsto \false} \\ 
\heap' = \heap, \pointer{p}{c} \mapsto o}
{\actconf{\actframe{\ell}{\actstate{onDestroy}}{\pi}{\ocallstack} :: \actstack}{\heap}{\sheap} \Rightarrow \actconf{\uactframe{\pointer{p}{c}}{\actstate{constructor}}{\pi}{\getcb{\pointer{p}{c}}{\actstate{constructor}}} :: \actstack}{H'}{\sheap}}

\inferrule[(A-Hidden)]
{\varphi = \actframe{\ell}{s}{\pi}{\ocallstack} \\
s \in \{\actstate{onResume},\actstate{onPause}\} \\
(s',s'') \in \{(\actstate{onPause},\actstate{onStop}),(\actstate{onStop},\actstate{onDestroy})\} }
{\actconf{\varphi :: \actstack :: \actframe{\ell'}{s'}{\pi'}{\ocallstack'} :: \actstack'}{\heap}{\sheap} \Rightarrow \actconf{\varphi :: \actstack :: \uactframe{\ell'}{s''}{\pi'}{\getcb{\ell'}{s''}} :: \actstack'}{\heap}{\sheap}}

\inferrule[(A-Start)]
{s \in \{\actstate{onPause},\actstate{onStop}\} \\
i = \intent{c}{(k \mapsto v)^*} \\
\emptyset \vdash \serblock{\heap}(i) = (i',\heap') \\
\pointer{p}{c},\pointer{p'}{\astart{c}} \not\in \dom(\heap,\heap') \\
o = \obj{c}{(f_{\tau} \mapsto \defvalue_{\tau})^*,\finished \mapsto \false, \fintent \mapsto \pointer{p'}{\astart{c}}, \parent \mapsto \ell} \\
\heap'' = \heap,\heap',\pointer{p}{c} \mapsto o, \pointer{p'}{\astart{c}} \mapsto i'}
{\actconf{\actframe{\ell}{s}{i :: \pi}{\ocallstack} :: \actstack}{\heap}{\sheap} \Rightarrow \actconf{\uactframe{\pointer{p}{c}}{\actstate{constructor}}{\varepsilon}{\getcb{\pointer{p}{c}}{\actstate{constructor}}} :: \actframe{\ell}{s}{\pi}{\ocallstack} :: \actstack}{\heap''}{\sheap}}

\inferrule*[width=30em,lab=(A-Swap)]
{\varphi' = \actframe{\ell'}{\actstate{onPause}}{\varepsilon}{\ocallstack'} \\
\heap(\ell').\finished = \true \\
\varphi = \actframe{\ell}{s}{i :: \pi}{\ocallstack} \\
s \in \{\actstate{onPause},\actstate{onStop}\} \\
\heap(\ell').\parent = \ell}
{\actconf{\varphi' :: \varphi :: \actstack}{\heap}{\sheap} \Rightarrow \actconf{\varphi :: \varphi' :: \actstack}{\heap}{\sheap}}

\inferrule*[width=50em,lab=(A-Result)]
{\varphi' = \actframe{\ell'}{\actstate{onPause}}{\varepsilon}{\ocallstack'} \\
\heap(\ell').\finished = \true \\
\varphi = \actframe{\ell}{s}{\varepsilon}{\ocallstack} \\
s \in \{\actstate{onPause},\actstate{onStop}\} \\
\heap(\ell').\parent = \ell \\
\emptyset \vdash \serval{\heap}(\heap(\ell').\result) = (\ell'',\heap') \\
\heap'' = (\heap,\heap')[\ell \mapsto \heap(\ell)[\result \mapsto \ell'']]}
{\actconf{\varphi' :: \varphi :: \actstack}{\heap}{\sheap} \Rightarrow \actconf{\uactframe{\ell}{s}{\varepsilon}{\getcb{\ell}{\actstate{onActivityResult}}} :: \varphi' :: \actstack}{\heap''}{\sheap}}
\end{mathpar}
where:
\begin{mathpar}
\inferrule
{}
{\serialized \vdash \serval{\heap}({\prim}) = (\prim, \cdot)}


\inferrule
{\pointer{p}{\ann} \in \serialized}
{\serialized \vdash \serval{\heap}(\pointer{p}{\ann}) = (\newpointer{\pointer{p}{\ann}}, \cdot)}

\inferrule
{\pointer{p}{\ann} \notin \serialized \\ 
\serialized \cup \{\pointer{p}{\ann}\} \vdash
\serblock{\heap}(\heap(\pointer{p}{\ann})) = (b, \heap'') \\
\heap' = \heap'',\newpointer{\pointer{p}{\ann}} \mapsto b }
{\serialized \vdash \serval{\heap}(\pointer{p}{\ann}) = (\newpointer{\pointer{p}{\ann}}, \heap')}

\inferrule
{\forall i \in [1,n]: \serialized \vdash \serval{\heap}(v_i) = {(u_i,\heap_i)} \\
\heap' = \heap_1, \ldots, \heap_n}
{\serialized \vdash \serblock{\heap}(\arr{\tau}{(v_i)^{i \leq n}}) = (\arr{\tau}{(u_i)^{i \leq n}}, \heap')}

\inferrule
{\forall i \in [1,n]: \serialized \vdash \serval{\heap}(v_i) = {(u_i,\heap_i)} \\
\heap' = \heap_1, \ldots, \heap_n}
{\serialized \vdash \serblock{\heap}(\obj{c'}{(f_i \mapsto v_i)^{i \leq n}}) = (\obj{c'}{(f_i \mapsto u_i)^{i \leq n}}, \heap')}

\inferrule
{\forall i \in [1,n]: \serialized \vdash \serval{\heap}(v_i) = {(u_i,\heap_i)} \\
\heap' = \heap_1, \ldots, \heap_n }
{\serialized \vdash \serblock{\heap}(\intent{c'}{(k_i \mapsto v_i)^{i \leq n}}) = (\intent{c'}{(k_i \mapsto u_i)^{i \leq n}}, \heap')}
\end{mathpar}
\textbf{Conventions:} the activity stack on the left-hand side does
not contain underlined frames, but for the first two rules. In the
serialization rules we assume the existence of a function $\newpointer{\_}$
assigning to each pointer a fresh pointer with the same annotation, used to 
store the result of the serialization.
\caption{Reduction Relation for Configurations ($\actconf{\actstack}{\heap}{\sheap} \Rightarrow \actconf{\actstack'}{\heap'}{\sheap'}$)}
\label{tab:activity-semantics}
\end{table*}

Rule \irule{A-Active}
allows the execution of the statements in the active frame, using the
reduction relation for local configurations described in
Section~\ref{sec:dalvik}. Rule \irule{A-Deactivate} models the
situation where the active frame has run up to completion: the frame
loses priority and one of the other rules can be applied. Rule
\irule{A-Step} models the transition of the top-level activity from
state $s$ to one of its successors $s'$ in the activity lifecycle:
correspondingly, a new callback method is executed. Two
side-conditions constrain the possible state transitions, based on the
presence of pending activities to start and on whether the activity has
finished or not. 

Rule \irule{A-Destroy} models the removal of a finished activity from the activity stack. Rule \irule{A-Back} models the scenario where the user hits the back button on the Android device and the top-most activity gets finished by the system. Rule \irule{A-Replace} corresponds to screen orientation changes: the foreground activity is destroyed and gets replaced by a fresh activity instance; notice that the new pointer to the heap is annotated with the class of the activity. Rule \irule{A-Hidden} models the scenario where a new activity (the frame $\varphi$) has come to the foreground and hides a previously running activity, which gets stopped or destroyed by the system.

The starting of a new activity is modelled by rule \irule{A-Start}. The top-most activity is paused or stopped and there is some intent $i$ to be sent to $c$: the intent is serialized and a new instance of $c$ is pushed on the activity stack, setting its $\fintent$ field to a pointer to the serialized copy of $i$ and setting its $\parent$ field to a pointer to the activity which sent the intent. The pointer to the new activity is annotated with the class $c$, while the pointer to the serialized copy of the intent gets the annotation $\astart{c}$: again, this is needed just for the static analysis and will be discussed later. Notice that, if multiple activities need to be started, rule \irule{A-Swap} allows a parent activity to substitute itself to a child activity on the top of the activity stack, so that rule \irule{A-Start} can be applied again to fire the remaining intents. Finally, rule \irule{A-Result} allows a finished activity in the foreground to return the result of its computation to the parent activity: the parent activity gets a serialized copy of the result and becomes active by executing a corresponding callback, bound to the \actstate{onActivityResult} state.

\subsection{Examples}
\label{sec:examples}
One reason why it is useful to have a formal semantics before devising
a static analysis technique is to pinpoint corner cases which may
potentially lead
to unsound analysis results. We discuss two examples below. 

\subsubsection{Static Fields}
Even though inter-component communication does not allow for the
exchange of references, activities in the same application can still
share memory by using static fields. This is apparent in the formal
semantics, since the syntax of configurations $\Psi$ contains a global
static heap $\sheap$, which can be accessed by using publicly known
names of static fields. We then observe that the order of execution of
different activities, or even different callbacks inside the same
activity, is very hard to predict: for instance, the rules in
Table~\ref{tab:activity-semantics} highlight that even activities
which are not on the top of the activity stack may become active and
execute callbacks by rule \irule{A-Hidden}. Also, the same callback
may be executed multiple times, since an activity may be routinely
recreated by the Android system due to user activities (e.g., screen
orientation changes), which cannot be known statically, 
as modelled by rule \irule{A-Replace}.  

The implication on static analysis is that it is extremely challenging to implement flow-sensitivity on accesses to static fields without producing unsound results. Furthermore, given that static fields may be used to share pointers to heap locations, flow-sensitivity for heap accesses is also hard to achieve. Since we target soundness in this work, the static analysis we devise in the next section is flow-insensitive on both static fields and heap locations.

\subsubsection{Serialization}
Rule \irule{A-Start} of the operational semantics highlights that
intents are serialized upon inter-component communication. This means
that, when a parent activity starts a child activity, the latter
operates on a copy of the intent sent by the former and not on the
same intent.  

The implication on static analysis is that, although the callback bound to the \actstate{onActivityResult} state of the parent activity is always executed after the construction of the child activity, no change to the intent done by the child activity should overwrite the original over-approximation of the intent computed for the parent
activity when a result is returned to it. This applies to any object which is serialized with the intent. The static analysis in the next section provides a conservative over-approximation of this behaviour.


\section{Static Analysis}
\label{sec:statics}
The static analysis we propose works by translating an input program $P$ into a corresponding \emph{abstract} program $\Delta$, i.e., a set of Horn clauses modelling an over-approximation of its semantics. By feeding these clauses to an automated theorem prover and by showing the unsatisfiability of an appropriate logical formula, we can prove that some set of undesired configurations is never reached by $P$.

\subsection{Overview}
The analysis is based on the syntactic categories in Table~\ref{tab:absdoms}. We start by discussing how values are approximated. We presuppose the existence of an arbitrary set of abstract domains used to approximate primitive values: for each primitive value $\prim$, we assume that there exists a corresponding abstraction $\absprim$, e.g., integer numbers could be approximated by their sign. Locations of the form $\ell = \pointer{p}{\ann}$, instead, are abstracted into their annotation $\ann$. An \emph{abstract value} $\absval$ is a set of elements drawn from either the abstract domains or the set of annotations.

\begin{table*}[htb]
\begin{mathpar}
\begin{array}{llcl}
\text{Facts} & \fact & \define & \\
\text{Abs. registers} & & | & \absreg{\spp}{t^*}{t^*} \\
\text{Abs. heap entries} & & | & \absheap(t,t') \\
\text{Abs. static fields} & & | & \abssheap_{\mathsf{c},\mathsf{f}}(t) \\
\text{Abs. right-hand sides} & & | & \prhs{t} \\
\text{Abs. results} & & | & \absresult{\mathsf{c},\mathsf{m}}{t^*}{t} \\
\text{Abs. pending activities} & & | & \absdispatch(t,t') \\
\text{Set membership} & & | & t \in t' \\
\text{Subtyping} & & | & t \subtype t' \\
\text{Horn clauses} & & | & \forall x^*.\bigwedge_i \fact_i \implies \fact \\
\text{Abs. programs} & \absprog & \define & \{\fact_1,\ldots,\fact_n\}
\end{array}
\begin{array}{llcl}
\text{Abs. values} & \absual,\absval & \define & \emptyset ~|~ \{\absprim\} ~|~ \{\absloc\} ~|~ \absval \cup \absval \\
\text{Abs. objects} & \hat{o} & \define & \absobj{c}{(f_{\tau} \mapsto \absval)^*} \\
\text{Abs. arrays} & \hat{a} & \define & \absarray{\tau}{\absval} \\
\text{Abs. intents} & \hat{i} & \define & \absintent{c}{\absval} \\
\text{Abs. mem. blocks} & \absblock & \define & \hat{o} ~|~ \hat{a} ~|~ \hat{i} \\
\\
\text{Variables} & x,y & \in & \textit{Vars} \\
\text{Constants} & \const{k} & \define & \absval ~|~ \absblock ~|~ \tau ~|~ \ann \\
\text{Terms} & t & \define & \const{k} ~|~ x ~|~ \astart{t}
\end{array}
\end{mathpar}
\caption{Abstract Domains and Analysis Facts}
\label{tab:absdoms}
\end{table*}

The different forms of annotations $\ann$ provide insight on different
aspects of the static analysis. Program point annotations $\pp =
c,m,\pc$ are used to represent pointers to memory blocks instantiated
using the statements \texttt{new}, \texttt{newarray} and
\texttt{newintent}: by abstracting these elements with the program
point where they are created, we implement a
\emph{plain-object-sensitive} static
analysis~\cite{Smaragdakis:2011:PYC:1926385.1926390}.  {We chose it because it is 
well-understood and  convenient to both formalize and
present: we plan to integrate more advanced analyses like 2full+1H in
future releases. Class name 
  annotations $c$, instead, are used to represent activities in an 
  object-insensitive way: different activities of the same class $c$ are all 
  abstracted by the annotation $c$, since it is generally hard to statically 
  discriminate between different activity instances. Finally, we use the annotation 
  $\astart{c}$ to abstract all the intents which are used to start an activity of class 
  $c$.

Coming to memory blocks, our analysis is field-sensitive on objects, but field-insensitive on both arrays and intents. It is easier to implement field-sensitivity for objects, since field names are statically known in Java. Implementing field-sensitivity for arrays would require precise information on array bounds and indexes; intents, instead, would need an accurate string analysis, to deal with their dictionary-like programming patterns. It would be possible to leverage existing proposals~\cite{DilligDA11} to implement a more precise analysis in terms of field-sensitivity, but we propose a simpler framework here to focus on the Android-specific aspects of the analysis. Notice that, just like the objects they approximate, abstract objects $\hat{o}$ feature type annotations on their fields, which are omitted when unimportant. 

Abstract values and abstract memory blocks, plus all the types available in the analysed program and the annotations, determine a universe of \emph{constants}, ranged over by $\const{k}$. A \emph{term} $t$ is either a constant $\const{k}$, a variable $x$ drawn from a denumerable set $\textit{Vars}$ disjoint from the set of constants, or an expression of the form $\astart{t'}$ for some term $t'$. The set of terms is used to define the syntax of \emph{facts} $\fact$, logical formulas built on selected predicate symbols used by the analysis.

The fact $\absreg{\apc}{\absual^*}{\absval^*}$ states that, whenever the method $m$ of class $c$ is invoked with some arguments over-approximated by $\absual^*$, the state of the local registers at the $\pc$-th statement is over-approximated by $\absval^*$. The syntax of the fact highlights that: (1) the analysis is flow-sensitive for register values, since it computes different static approximations at different program points, and (2) method invocations are handled in a context-sensitive way, where the notion of context coincides with the (abstraction of) the actual arguments supplied to the method upon invocation. The fact $\absheap(\ann,\absblock)$ states that some location $\pointer{p}{\ann}$ refers to a heap element storing a memory block over-approximated by $\absblock$ at some point of the program execution. Notice that the fact does not contain any program point information, i.e., the analysis is flow-insensitive for heap locations, which is important for soundness (see Section~\ref{sec:examples}). Similarly, the fact $\abssheap_{\mathsf{c},\mathsf{f}}(\absval)$ states that the static field $f$ of class $c$ contains a value which is over-approximated by $\absval$ at some point of the program execution. The fact $\prhs{\absval}$ states that the right-hand side of the \texttt{move} statement at program point $\pp$ evaluates to a value over-approximated by $\absval$. The fact $\absresult{\mathsf{c},\mathsf{m}}{\absual^*}{\absval}$ states that, whenever the method $m$ of class $c$ is invoked with some arguments over-approximated by $\absual^*$, its return value is over-approximated by $\absval$. The fact $\absdispatch(c,\hat{i})$ tracks that an activity of class $c$ has sent an intent which is over-approximated by $\hat{i}$. We then have set membership facts $t \in t'$ and subtyping facts $\tau \subtype \tau'$ with the obvious meaning. 

Finally, Horn clauses define the abstract semantics of programs. A Horn clause has the form:
\[ 
\forall x_1,\ldots,\forall x_m.\fact_1 \wedge \ldots \wedge \fact_n \implies \fact,
\]
where all the variables of $\fact_1,\ldots,\fact_n,\fact$ belong to $\{x_1,\ldots,x_m\}$ and each variable of $\fact$ occurs among the variables of $\fact_1,\ldots,\fact_n$. Since most of the Horn clauses we present do not make use of constants, to improve readability we omit the universal quantifiers in front of Horn clauses and we just represent each variable occurring therein with a constant of the expected type. The few exceptions where constants are actually used are disambiguated using a $\mathsf{sans\ serif}$ font, e.g., we use $\const{c}$ to denote the constant corresponding to the activity class $c$ specifically, rather than some universally quantified variable standing for an arbitrary activity class. We let an underscore (\_) stand for any syntactic element occurring in a Horn clause which is not significant to understanding.

\subsection{Analysis Specification}

\subsubsection{Abstract Semantics of Dalvik}
We start by presenting the abstract evaluation rules for right-hand sides, which are simple and provide a good intuition on how the static analysis works. These rules are given in Table~\ref{tab:abs-rhs}.

\begin{table*}[htb]
\begin{mathpar}
\arhs{\prim} = \{\prhs{\{\absprim\}}\}

\arhs{r_i} = \{\absreg{\spp}{ \_ }{\absval^*} \implies \prhs{\absval_i }\}

\arhs{c.f} = \{\abssheap_{\mathsf{c},\mathsf{f}}(\absval) \implies \prhs{\absval}\}

\arhs{r_i.f} = \{\absreg{\spp}{ \_ }{\absval^*} \wedge \absloc \in \absval_i \wedge \absheap(\absloc,\absobj{c}{(f' \mapsto \absval')^*, f \mapsto \absual}) \implies \prhs{\absual}\}

\arhs{r_i[r_j]} = \{\absreg{\spp}{ \_ }{\absval^*} \wedge \absloc \in \absval_i \wedge \absheap(\absloc, \absarray{\tau}{\absual}) \implies \prhs{\absual}\}
\end{mathpar}
\caption{Abstract Evaluation of Right-hand Sides}
\label{tab:abs-rhs}
\end{table*}

To abstract a primitive value $\prim$ at any program point $\pp$, we just pick the corresponding element $\absprim$ from the underlying abstract domain. To abstract the content of the register $r_i$ at program point $\pp$, we take the fact $\absreg{\spp}{ \_ }{\absval^*}$ and we return the $i$-th abstract value $\absval_i$. To abstract the content of a static field $c.f$ at any program point, we take any fact $\abssheap_{\mathsf{c},\mathsf{f}}(\absval)$ and we return the abstract value $\absval$. Abstracting the content of the field $f$ of an object at program point $\pp$ is slightly more complicated: if the pointer to the object is stored in the register $r_i$, we pick the $i$-th abstract value $\absval_i$ from the fact $\absreg{\spp}{ \_ }{\absval^*}$ modelling the state of the registers at $\pp$; then, if $\absval_i$ contains any pointer abstraction $\ann$, we use it to match a corresponding abstract heap entry $\absheap(\ann,\hat{o})$ and we return the value of the field $f$ of the abstract object $\hat{o}$ contained therein. We similarly abstract the content of array cells: just notice that, since the representation of arrays is field-insensitive, the index of the cell does not play any role in the static analysis.

The rules for abstracting a right-hand side are useful to define the abstract semantics of the \texttt{move} statement. Other statements require some additional definitions. First, for each comparison operator $\comp$ and each primitive operation $\odot,\oplus$ of the concrete semantics, we presuppose the existence of a corresponding abstract operation $\acomp$, $\aunop$ and $\abinop$ defined over the elements of the appropriate abstract domain. Then, given an abstract memory block $\hat{b}$, we define a function $\absgettype(\hat{b})$ as follows:
\[
\absgettype(\hat{b}) =
\begin{cases}
c & \text{if } \hat{b} = \absobj{c}{(f \mapsto \absval)^*} \\
\arrtype{\tau} & \text{if } \hat{b} = \absarray{\tau}{\absval} \\
\texttt{Intent} & \text{if } \hat{b} = \absintent{c}{\absval}
\end{cases}
\]
Finally, we assume a function $\abslookup(m)$, which returns the set of classes which define (or inherit) a method called $m$.

With these definitions, we are ready to introduce the abstract semantics of statements. The idea is to define, for each possible form of statement $\stm$, a translation $\ainst{\stm}$ into a set of Horn clauses, which over-approximate the semantics of $\stm$ at program point $\pp$. The full formal semantics of the translation is given in Table~\ref{tab:abs-statements} and explained below. 

\begin{table*}[htb] \small
\begin{mathpar}
\begin{array}{lcl}
\ainst{\goto \pc'} & = & \{\absreg{\spp}{ \_ }{\absval^*} \implies \absreg{\apcp}{ \_ }{\absval^*}\} \\

\ainst{\ifbr {r_i}{r_j} {\pc'}} & = & \{\absreg{\spp}{ \_ }{\absval^*} \wedge \absval_i\ \acomp\ \absval_j \implies \absreg{\apcp}{ \_ }{\absval^*}\}\, \cup \\
& & \{\absreg{\spp}{ \_ }{\absval^*} \wedge \neg(\absval_i\ \acomp\ \absval_j) \implies \absreg{\apcn}{ \_ }{\absval^*}\} \\

\ainst{\binop{r_d}{r_i}{r_j}} & = & \{\absreg{\spp}{ \_ }{\absval^*} \implies
\absreg{\apcn}{ \_ }{\absval^*[d \mapsto \absval_i\ \abinop\ \absval_j]}\} \\

\ainst{\unop{r_d}{r_i}} & = & \{\absreg{\spp}{ \_ }{\absval^*} \implies
\absreg{\apcn}{ \_ }{\absval^*[d \mapsto \aunop\, \absval_i]}\} \\

\ainst{\move{r_d}{\rhs}} & = & \{\prhs{\absval'} \wedge \absreg{\spp}{\_}{\absval^*} \implies \absreg{\apcn}{ \_} {\absval^*[d \mapsto \absval']}\} \cup \arhs{\rhs} \\

\ainst{\move{r_a[r_\ind]}{\rhs}} & = & \{\prhs{\absval''} \wedge \absreg{\spp}{\_ }{\absval^*} \wedge \absloc \in \absval_a \wedge \absheap(\absloc,\absarray{\tau}{\absval'}) \implies \absheap(\absloc,\absarray{\tau}{\absval' \cup \absval''})\}\, \cup \\
& &  \{\absreg{\spp}{ \_ }{\absval^*} \implies \absreg{\apcn}{ \_ }{\absval^*}\} \cup \arhs{\rhs} \\

\ainst{\move{r_o.f}{\rhs}} & = & \{\prhs{\absval''} \wedge \absreg{\spp}{ \_} {\absval^*} \wedge \absloc \in \absval_o \wedge \absheap(\absloc,\absobj{c'}{(f' \mapsto \absual')^*, f \mapsto \absval'}) \implies \\
& & \absheap(\absloc,\absobj{c'}{(f' \mapsto \absual')^*, f \mapsto \absval'')})\} \cup \{\absreg{\spp}{ \_} {\absval^*} \implies \absreg{\apcn}{ \_ }{\absval^*}\} \cup \arhs{\rhs} \\

\ainst{\move{c'.f}{\rhs}} & = & \{\prhs{\absval'} \implies \abssheap_{\mathsf{c'},\mathsf{f}}(\absval')\} \cup \{\absreg{\spp}{ \_ }{\absval^*} \implies \absreg{\apcn}{ \_}{\absval^*}\} \cup \arhs{\rhs} \\

\ainst{\instanceof{r_d}{r_s}{\tau}} & = & \{\absreg{\spp}{ \_}{\absval^*} \wedge \absloc \in \absval_s \wedge \absheap(\absloc, \hat{b}) \wedge \absgettype(\hat{b}) \subtype \tau \implies \absreg{\apcn}{ \_ }{\absval^*[d \mapsto \widehat{\true}]}\}\, \cup \\
& & \{\absreg{\spp}{ \_ }{\absval^*} \wedge \absloc \in \absval_s \wedge \absheap(\absloc,\hat{b}) \wedge \absgettype(\hat{b}) \not\subtype \tau \implies \absreg{\apcn}{ \_ }{\absval^*[d \mapsto \widehat{\false}]}\} \\

\ainst{\checkcast{r_s}{\tau}} & = & \{\absreg{\spp}{ \_ }{\absval^*} \wedge \absloc \in \absval_s \wedge \absheap(\absloc,\hat{b}) \wedge \absgettype(\hat{b}) \subtype \tau \implies \absreg{\apcn}{ \_}{\absval^*}\} \\

\ainst{\invoke{r_o}{m'}{(r_{i_j})^{j \leq n}}} & = & \{\absreg{\spp}{ \_ }{\absval^*} \wedge \absloc \in \absval_o \wedge \absheap(\absloc,\absobj{c'}{(f \mapsto \absual)^*}) \wedge c' \subtype \const{c''} \implies \\
& & \absreg{\mathsf{c''},\mathsf{m'},\mathsf{0}}{(\absval_{i_j})^{j \leq n}}{(\adefvalue_k)^{k \leq \loc}, (\absval_{i_j})^{j \leq n}}  ~|~ c'' \in \abslookup(m') \wedge \sign(c'',m') = \methsign{(\tau_j)^{j \leq n}}{\tau}{\loc}\}\, \cup \\
& & \{ \absreg{\spp}{ \_ }{\absval^*} \wedge \absloc \in \absval_o \wedge \absheap(\absloc,\absobj{c'}{(f \mapsto \absual)^*}) \wedge c' \subtype \const{c''} \wedge \absresult{\mathsf{c''},\mathsf{m'}}{(\absval_{i_j})^{j \leq n}}{\absval'_{\ret}} \implies \\
& & \absreg{\apcn}{ \_} {\absval^*[\ret \mapsto \absval'_{\ret}]} ~|~ c'' \in \abslookup(m')\} \\

\ainst{\sinvoke{c'}{m'}{(r_{i_j})^{j \leq n}}} & = & \{\absreg{\spp}{ \_ }{\absval^*} \implies \absreg{\mathsf{c'},\mathsf{m'},\mathsf{0}}{(\absval_{i_j})^{j \leq n}}{(\adefvalue_k)^{k \leq \loc},(\absval_{i_j})^{j \leq n}} ~|~ \sign(c', m') = \methsign{(\tau_j)^{j \leq n}}{\tau}{\loc} \}\, \cup \\
& & \{\absreg{\spp}{ \_ }{\absval^*} \wedge \absresult{\mathsf{c'},\mathsf{m'}}{(\absval_{i_j})^{j \leq n}}{\absval'_{\ret}} \implies \absreg{\apcn}{ \_} {\absval^*[\ret \mapsto \absval'_{\ret}]}\} \\

\ainst{\new{r_d}{c'}} & = & \{ \absreg{\spp}{\_}{\absval^*} \implies \absheap(\spp, \absobj{c'}{(f \mapsto \adefvalue_{\tau})^*}\} \cup \{ \absreg{\spp}{ \_ }{\absval^*} \implies \absreg{\apcn}{\_}{\absval^*[d \mapsto \spp]}\} \\

\ainst{\newarray{r_d}{r_l}{\tau}} & = & \{ \absreg{\spp}{\_}{\absval^*} \implies \absheap(\spp, \absarray{\tau}{\adefvalue_{\tau}})\} \cup \{ \absreg{\spp}{ \_ }{\absval^*} \implies \absreg{\apcn}{\_}{\absval^*[d \mapsto \spp]}\} \\

\ainst{\return} & = &  \{\absreg{\spp}{\absval^*_{call}}{\absval^*} \implies \absresult{\mathsf{c},\mathsf{m}}{\absval^*_{call}}{\absval_{\ret}}\} \\

\ainst{\startact{r_i}} & = & \{\absreg{\spp}{\_}{\absval^*} \wedge \absloc \in \absval_i \wedge \absheap(\absloc,\absintent{c'}{\absual}) \implies \absdispatch(\const{c},\absintent{c'}{\absual})\}\, \cup \\ 
& & \{ \absreg{\spp}{\_ }{\absval^*} \implies \absreg{\apcn}{\_}{\absval^*}\} \\

\ainst{\newintent{r_d}{c'}} & = & \{\absreg{\spp}{\_}{\absval^*} \implies \absheap(\spp, \absintent{c'}{\emptyset})\} \cup \{\absreg{\spp}{ \_ }{\absval^*} \implies \absreg{\apcn}{\_}{\absval^*[d \mapsto \spp]}\} \\

\ainst{\putextra{r_i}{r_k}{r_j}} & = & \{\absreg{\spp}{\_}{\absval^*} \wedge \absloc \in \absval_i \wedge \absheap(\absloc, \absintent{c'}{\absval'}) \implies \absheap(\absloc, \absintent{c'}{\absval' \cup \absval_j})\}\, \cup  \\
& & \{\absreg{\spp}{ \_ }{\absval^*} \implies \absreg{\apcn}{\_}{\absval^*}\} \\

\ainst{\getextra{r_i}{r_k}{\tau}} & = & \{\absreg{\spp}{\_}{\absval^*} \wedge \absloc \in \absval_i \wedge \absheap(\absloc, \absintent{c'}{\absval'})  \implies \absreg{\apcn}{\_}{\absval^*[\ret \mapsto \absval']}\}
\end{array}
\end{mathpar}
\caption{Abstract Semantics of \sem{} - Statements (let $\pp = c,m,\pc$)}
\label{tab:abs-statements}
\end{table*}

The rule for $\goto{\pc'}$ propagates the state of the registers at the current program counter $\pc$ to $\pc'$. The rule for $\ifbr{r_i}{r_j}{\pc'}$ propagates the state of the registers at the current program counter $\pc$ either to $\pc'$ or to $\pc+1$, based on the outcome of a comparison $\acomp$ between the abstract values $\absval_i$ and $\absval_j$ approximating the content of registers $r_i$ and $r_j$ respectively: both branches may be enabled, as the result of an over-approximation of the contents of the registers. The two rules for unary and binary operations just employ the appropriate abstract operation to update the approximation of the content of the destination register $r_d$. The four rules for the \texttt{move} statement rely on the auxiliary rules for abstracting a right-hand side we introduced before: these rules store their result in a $\mathsf{RHS}$ fact, which occurs in the premises of the Horn clause used to update the abstraction of the left-hand side. The most interesting point to notice here is that field-sensitivity or its absence has an import on how fields are updated: for objects, we replace the old value of the field with the new one; for arrays and intents, instead, we add the new value to the old approximation, since their abstraction over-approximates the content of the entire data structure, rather than just the single element which is updated. The rules for \texttt{instof} and \texttt{checkcast} use the $\absgettype$ function previously defined.

The rule for \texttt{invoke} is the most complicated one, since it has to deal with dynamic dispatching. The challenge here is that the name of the invoked method is statically known from the syntax of the statement, but the method implementation is not, since it depends on the runtime type of the receiver object, an information which is only over-approximated when solving the Horn clauses, rather than when generating them. We then use the method name and the number of arguments passed upon invocation to narrow the set of possible classes of the receiver object, using the functions $\abslookup$ and $\sign$, and we generate one Horn clause for each of them. We then rely on subtyping to make the analysis precise, by imposing that a Horn clause generated for class $c''$ can only be fired if the class $c'$ of (the abstraction of) the receiver object is a subtype of $c''$. Besides implementing a sound approximation of the dynamic dispatching mechanism, the rule for \texttt{invoke} generates additional Horn clauses used to propagate the abstraction of the method return value from the callee to the caller: this is done by using a $\mathsf{Res}$ fact, which is introduced by a \texttt{return} statement in the implementation of the callee, as we discuss below. The rule for static method invocation follows a similar logic, but it is significantly simpler, due to the lack of dynamic dispatching on static calls.

The rules for object and array creation create a new abstract heap entry $\absheap(\ann,\absblock)$, where $\ann$ is the current program point and $\absblock$ is the abstraction of a freshly initialized object/array. The rule for \texttt{return} introduces a $\mathsf{Res}$ fact, storing an over-approximation of the method return value; notice that the arguments $\absval_{call}^*$ supplied upon method invocation are propagated in the $\mathsf{Res}$ fact, which is important to implement context-sensitivity, i.e., to propagate the result to the right caller. The rule for \texttt{start-activity} tracks that the present activity $c$ has sent an intent: an over-approximation of the intent is propagated from the corresponding abstract heap entry into the $\mathsf{I}$ fact modelling the presence of a pending activity which is about to start. The last rules for managing intents should be easy to understand, based on the intuitions given for the other rules.

\subsubsection{Abstract Semantics of Activities}
We can finally introduce the abstract semantics of activities. Intuitively, it is defined by: (1) the Horn clauses produced by translating each statement in the bytecode, and (2) a small set of bytecode-independent Horn clauses, abstracting the event-driven behaviour of activities. This is formalized next.

\begin{definition}
Let $P = (\class_i)^{i \leq n}$ be a program where $\class_i = \cls{c_i}{c'}{c^*}{\field^*}{(\method_j)^{j \leq h_i}}$ and $\method_j = \meth{m_j}{\methsign{\tau^*}{\tau}{\loc}}{(\stm_k)^{k \leq s_{ij}}}$, we let $\translate{P}$ be defined as follows:
\[
\translate{P} = \bigcup_{i \leq n, j \leq h_i, k \leq s_{ij}} \ainstfull{\stm_k}{c_i,m_j,k} \cup \mathcal{R},
\]
where $\mathcal{R}$ stands for the union of all the rules in Table~\ref{tab:abs-activity}.
\end{definition}

\begin{table*}[htb]
\begin{mathpar}
\begin{array}{lcl}
\rulename{Cbk} & = & \{\absheap(c,\absobj{c}{(f \mapsto \_)^*}) \wedge c \subtype \const{c'} \implies \absreg{\mathsf{c'},\mathsf{m},\mathsf{0}}{(\top_{\tau_j})^{j \leq n}}{(\adefvalue_k)^{k \leq \loc},c,(\top_{\tau_j})^{j \leq n}} ~|~ \\
& & c' \text{ is an activity class} \wedge \exists s: m \in \cb(c',s) \wedge \sign(c',m) = \methsign{\tau_1,\ldots,\tau_n}{\tau}{\loc}\} \\

\rulename{Fin} & = & \{\absheap(c,\absobj{c}{(f \mapsto \_)^*,\finished \mapsto \_}) \implies \absheap(c,\absobj{c}{(f \mapsto \_)^*,\finished \mapsto \top_{\type{bool}}})\} \\

\rulename{Rep} & = & \{\absheap(c,\absobj{c}{(f_{\tau} \mapsto \_)^*}) \implies  \absheap(c,\absobj{c}{(f_{\tau} \mapsto \adefvalue_{\tau})^*})\} \\

\rulename{Act} & = & \{\absdispatch(c',\absintent{c}{\absval})) \implies \absheap(\astart{c},\absintent{c}{\absval})\}\, \cup \\
& & \{\absdispatch(c',\absintent{c}{\absval})) \implies \absheap(c, \absobj{c}{(f_{\tau} \mapsto \adefvalue_{\tau})^*, \finished \mapsto \widehat{\false}, \parent \mapsto c', \fintent \mapsto \astart{c}})\} \\

\rulename{Res} & = & \{\absheap(c',\absobj{c'}{(f' \mapsto \_)^*,\parent \mapsto c,\result \mapsto \absloc} \wedge \absheap(c,\absobj{c}{(f \mapsto \_)^*,\result \mapsto \_} \implies \\
& & \absheap(c,\absobj{c}{(f \mapsto \_)^*,\result \mapsto \absloc}\} \\

\rulename{Sub} & = & \{\tau \subtype \tau' ~|~ \tau \subtype \tau' \text{ is a valid subtyping judgement} \}
\end{array}
\end{mathpar}
\caption{Abstract Semantics of \sem{} - Activity Rules}
\label{tab:abs-activity}
\end{table*}

We explain the rules from Table~\ref{tab:abs-activity}. Rule \rulename{Cbk} simulates the invocation of a callback: since we do not approximate the activity state in the abstract semantics, any callback method bound to a state $s$ of the activity lifecycle may be non-deterministically dispatched; the statically unknown arguments supplied to the callback are abstracted by the top element ($\top$) of the abstract domain associated to their type, which is a sound over-approximation of any value of that type. Rule \rulename{Fin} tracks updates to the $\finished$ field of an activity in the abstract semantics: since it is hard to statically track whether an activity has finished or not, the rule sets the field to the top element of the abstract domain used to represent boolean value ($\top_{\type{bool}}$). Rule \rulename{Rep} approximates the behaviour of rule \irule{A-Replace} of the concrete semantics: the activity fields may be reset to their default abstract value as the result of a screen orientation change. 

Rule \rulename{Act} represents the starting of a new activity. If an intent has been sent by an activity of class $c'$ to start an activity of class $c$, we introduce: (1) a new abstract heap entry to bind an abstraction of the intent to $\astart{c}$, and (2) a new abstract heap entry to bind an abstraction of the started activity to $c$. No serialization happens in the abstract semantics: if an intent is used to send an object in the concrete semantics, a reference to the corresponding abstract object is sent in our abstraction. This is sound, since our analysis is flow-insensitive on heap values, hence no over-approximation of the original object is ever lost as the result of an update to the heap at the receiver side. We then have rule \rulename{Res}, which is used to communicate a result from a child activity to its parent, thus simulating the behaviour of rule \irule{A-Result} in the concrete semantics; again, no serialization happens in the process, rather a pointer to the result is passed. Finally, rule \rulename{Sub} corresponds to an axiomatization of the subtyping relationships for the analysed program.

\subsection{Formal Results}
The soundness of the analysis is proved using \emph{representation functions}, a standard approach in program analysis~\cite{NielsonNH99}.
The representation function $\rfconf$ maps an arbitrary configuration $\Psi$ into a corresponding set of facts $\absprog$, modelling an over-approximation of $\Psi$. Its definition is lengthy, but unsurprising, e.g., each element $\ell \mapsto b$ of the heap is converted into an abstract heap entry $\absheap(\ann,\absblock)$, where $\ann$ is the annotation on $\ell$ and $\absblock$ is an abstraction of $b$. After defining $\rfconf$, we introduce a partial order $\sqsubseteq$ on analysis facts, with the intuitive understanding that $\fact \sqsubseteq \fact'$ whenever $\fact$ is a more precise abstraction than $\fact'$. The partial order is then lifted to abstract programs by having
$\absprog <: \absprog'$ if and only if $\forall \fact \in \absprog: \exists \fact' \in \absprog': \fact \sqsubseteq \fact'$.

Our main theorem states that any reachable configuration in the concrete semantics is over-approximated by some set of facts which is provable using the abstract semantics of the program and an abstraction of the initial configuration. The proof is parametric with respect to the choice of the abstract domains/operations used for primitive values, provided they offer some minimal soundness guarantees. This allows for choosing different trade-off between efficiency and precision of the analysis.

\begin{theorem}[Preservation]
If $\Psi \Rightarrow^* \Psi'$ under a program $P$, there exists $\absprog :> \rfconf(\Psi')$ such that:
\[
\translate{P} \cup \rfconf(\Psi) \vdash \absprog.
\]
\end{theorem}

By providing an over-approximation of any reachable configuration of the concrete semantics in terms of a corresponding set of facts, the theorem can be used to prove the absence of undesired information flows of sensitive data into local registers of selected sink methods. In particular, we leverage the theorem to develop a provably sound taint analysis, based on standard ideas. Due to space constraints, we  refer to \iffull Appendix~\ref{sec:proofs} \else the online version~\cite{full} \fi for full details. 


\section{Experiments}
\label{sec:experiments}
We developed \tool, a static analysis tool for Android applications based on our theory.
\tool{} implements a sound, fully automatic taint analysis aimed at detecting malicious information flows in Android applications. The analysis is based on a publicly available database of sources and sinks specific to the Android platform~\cite{RasthoferAB14}.

\begin{figure}[htb]
\includegraphics[scale=0.27]{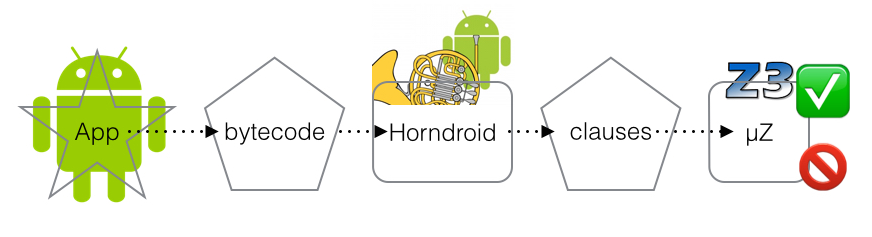}
\caption{\tool{} Architecture}
\label{fig:architecture}
\end{figure}

The architecture of \tool{} is shown in Figure~\ref{fig:architecture}. Given an Android application as an input, \tool{} generates Horn clauses defining an over-approximation of the application semantics, following the formal specification in Section~\ref{sec:statics}; the choice of the underlying abstract domains and operations implements a simple taint propagation logic. The Horn clauses are encoded in the SMT-LIB format supported by many popular SMT solvers, including our choice Z3~\cite{MouraB08}. \tool{} automatically generates analysis queries based on its database of sources and sinks\footnote{We use the latest and largest database available in the literature, i.e. the one used in DroidSafe~\cite{GordonKPGNR15}.} and the unsatisfiability of the queries is verified using the Property-Directed Reachability (PDR) engine implemented in Z3~\cite{HoderB12}. If no query is satisfiable, no information leak from a source to a sink may occur in the analysed application.

\subsection{Evaluation on DroidBench}
DroidBench~\cite{ArztRFBBKTOM14} is a set of small applications which
has been proposed by the research community as a testing ground for
static information flow analysis tools for Android. The current
version of the benchmark (2.0) includes 120 test cases, featuring both
leaky (positive) and benign (negative) examples. We tested IccTA, AmanDroid, DroidSafe and \tool{} on this benchmark, the results are summarized in the confusion matrix in Table~\ref{tab:droidbench}, reporting the number of true positives ($tp$), true negatives ($tn$), false positives ($fp$) and false negatives ($fn$) produced by the tools.

\begin{table}[htb]
\begin{center}
\begin{tabular}{c|c|c|}
\cline{2-3} & \multicolumn{2}{ c| }{Output} \\ 
\cline{2-3} & \emph{leaky} & \emph{benign} \\ 
\cline{2-3}
\cline{2-3} & \emph{IccTA}/\emph{AD}/\emph{DS}/\emph{HD} & \emph{IccTA}/\emph{AD}/\emph{DS}/\emph{HD} \\
\cline{1-3}
\multicolumn{1}{ |c| }{\emph{leaky}} & $tp:$ 64 / 70 / 89 / 96 & $fn:$ 36 / 30 / 11 / 4 \\ 
\cline{1-3}
\multicolumn{1}{ |c| }{\emph{benign}} & $fp:$ 8 / 5 / 10 / 6 & $tn:$ 11 / 14 / 9 / 13 \\
\cline{1-3}
\end{tabular}
\end{center}
\caption{Confusion Matrix on DroidBench}
\label{tab:droidbench}
\end{table}

IccTA does not detect 36 out of 100 leaky applications, AmanDroid misses 30 and DroidSafe still misses 11. Most of the leaks missed by IccTA and AmanDroid are due to flow-sensitivity and some callbacks which are not correctly detected by the analysis; as to DroidSafe, we do not have definite answers on the unsound results, given the sheer size of the project and the lack of complete documentation. \tool{} performs much better than all its competitors on DroidBench, since it only misses 4 leaky applications: all these cases are related to \emph{implicit flows}, which are not covered by standard taint analyses (and our formal proof).  

But even better, despite the strong security guarantees it provides,
the analysis performed by \tool{} is not overly conservative, since it
detects as potentially leaky only 6 out of 19 benign applications. We
notice that  3 of these false alarms are due to  flow insensitivity of
the heap abstraction, one to an over-approximation of exceptions, and
2 to an over-approximated treatment of inter-app communication. Only
AmanDroid is more precise, since it produces one less false positive;
on the other hand, it misses many more malicious information flows
than \tool{} (30 vs 4). For the sake of completeness, we report in Table~\ref{tab:droidbench-res}
a full breakdown of the experiments on DroidBench, omitting the cases where all the tools agree with the ground truth.

The experimental results on DroidBench are summarized by a few standard statistical measures in Table~\ref{tab:spec-sens}, which highlight that soundness in HornDroid does not come at the
cost of precision.

\begin{table}[htb]
\begin{center}
\begin{tabular}{c|c|c|c|c|}
\cline{2-5}
& \emph{IccTA} & \emph{AD} & \emph{DS} & \emph{HD} \\
\cline{1-5}
\multicolumn{1}{|c|}{\emph{Sensitivity}} & 0.64 & 0.70 & 0.89 & 0.96 \\ 
\cline{1-5}
\multicolumn{1}{|c|}{\emph{Specificity}} & 0.58 & 0.74 & 0.47 & 0.68 \\
\cline{1-5}
\multicolumn{1}{|c|}{\emph{F-Measure}} & 0.61 & 0.72 & 0.62 & 0.80 \\
\cline{1-5}
\end{tabular}
\end{center}
\emph{Sensitivity} = $tp / (tp + fn)$ $\sim$ Soundness \\
\emph{Specificity} = $tn / (tn + fp)$ $\sim$ Precision \\
\emph{F-Measure} = $2 * (sens * spec) / (sens + spec)$ $\sim$ Aggregate
\caption{Performance Measures on DroidBench}
\label{tab:spec-sens}
\end{table}

\iffull
\begin{table*}\small
\centering
\begin{tabular}{lllllll}
\emph{Category} & \emph{Case} & \emph{Leaky?} & \emph{IccTA} & \emph{DS} & \emph{AD} & \emph{HD} \\
\hline
Aliasing & Merge1 & \nt & \yf & \yf & \nt & \yf \\
\hline
Android-Specific 
&Parcel1 & \yt & \nf & \yt & \yt & \yt \\
&PublicAPIField1 & \yt & \nf  & \yt & \nf & \yt \\
&PublicAPIField2 & \yt & \nf & \yt & \nf & \yt \\
\hline
Arrays and Lists &ArrayAccess1 & \nt & \yf & \yf & \yf & \nt \\
&ArrayAccess2 & \nt & \yf & \yf & \yf & \nt \\
&ArrayCopy1 & \yt & \yt  & \yt & \nf & \yt \\
&ArrayToString1 & \yt & \nf & \yt & \yt  & \yt \\
&HashMapAccess1 & \nt & \yf & \yf & \nt  & \nt \\
&ListAccess1 & \nt & \yf & \yf & \yf & \nt \\
&MultidimensionalArray1 & \yt & \yt & \nf & \yt & \yt \\
\hline
Callbacks  
&MultiHandlers1 & \yt & \nf & \nf & \nf & \yt \\
&Ordering1 & \yt & \nf & \yt & \yt & \yt \\
&RegisterGlobal1 & \yt & \yt & \yt & \nf & \yt \\
&RegisterGlobal2 & \yt & \yt & \yt & \nf & \yt \\
&Unregister1 & \nt & \yf & \yf & \yf & \yf \\
\hline
Emulator Detection & ContentProvider1 & \yt & \yt & \yt & \nf & \yt \\
& IMEI1 & \yt & \nf & \nf & \nf & \nf \\
& PlayStore1 & \yt & \yt & \yt & \nf & \yt \\
\hline
Fields and Object Sensitivity  
&FieldSensitivity4 & \nt & \nt & \yf & \nt & \yf \\
&ObjectSensitivity2 & \nt & \nt & \yf & \nt & \yf \\
\hline
General Java 
& Exceptions3 & \nt & \yf & \yf & \yf & \yf \\
&Serialization1 & \yt & \nf & \yt & \nf & \yt \\
&StartProcessWithSecret1 & \yt & \nf & \yt & \nf & \yt \\
&StaticInitialization1 & \yt & \nf & \yt & \yt & \yt \\
&StaticInitialization3 & \yt & \nf & \yt & \yt & \yt  \\
&StringFormatter1 & \yt & \nf  & \yt & \nf & \yt \\
&StringPatternMatching1 & \yt & \nf & \yt & \yt & \yt \\
&StringToCharArray1 & \yt & \yt & \yt & \nf & \yt \\
&StringToOutputStream1 & \yt & \nf & \yt & \yt & \yt \\
&VirtualDispatch3 & \nt & \yf & \nt & \nt & \nt \\
\hline
Implicit Flows & ImplicitFlow1 & \yt & \nf & \yt & \nf & \yt \\
& ImplicitFlow2 & \yt & \nf & \nf & \nf & \nf \\
& ImplicitFlow3 & \yt & \nf & \nf & \nf & \nf \\
& ImplicitFlow4 & \yt & \nf & \nf & \nf & \nf \\
\hline
Inter-App Communication & Echoer & \yt & \nf & \yt & \nf & \yt \\
& SendSMS & \yt & \yt & \yt & \nf & \yt \\
\hline
Inter-Component Communication &ActivityCommunication1 & \yt & \yt & \yt & \nf & \yt \\
&ActivityCommunication3 & \yt & \nf & \yt & \yt & \yt \\
&ActivityCommunication6 & \yt & \nf & \yt & \yt & \yt \\
&ComponentNotInManifest1 & \nt & \nt & \yf & \nt & \yf \\
&IntentSink1 & \yt & \nf & \yt & \yt & \yt \\
&IntentSink2 & \yt & \nf & \yt & \nf & \yt  \\
&IntentSource1 & \yt & \nf & \yt & \yt & \yt  \\
&ServiceCommunication1 & \yt & \nf & \yt & \yt & \yt \\
&Singletons1 & \yt & \nf & \nf & \nf & \yt \\
\hline
Lifecycle &ActivityLifecycle1 & \yt & \nf & \yt & \yt & \yt \\
&ActivitySavedState1 & \yt & \nf & \yt & \yt & \yt \\
&ApplicationLifecycle1 & \yt & \yt & \yt & \nf & \yt \\
&ApplicationLifecycle2 & \yt & \yt & \yt & \nf & \yt \\
&ApplicationLifecycle3 & \yt & \yt & \yt & \nf & \yt \\
&BroadcastReceiverLifecycle2 & \yt & \nf & \yt & \nf & \yt \\
&FragmentLifecycle1 & \yt & \nf & \yt & \yt & \yt \\
&FragmentLifecycle2 & \yt & \nf & \yt & \nf & \yt \\
&SharedPreferenceChanged1 & \yt & \yt & \nf & \yt & \yt \\
\hline
Reflection & Reflection1 & \yt & \yt & \nf & \yt & \yt \\
& Reflection2 & \yt & \nf & \nf & \nf & \yt \\
& Reflection3 & \yt & \nf & \yt & \yt & \yt \\
& Reflection4 & \yt & \nf & \nf & \yt & \yt \\
\hline
Threading 
& Executor1 & \yt & \yt & \yt & \nf & \yt \\
& JavaThread1 & \yt & \yt & \yt & \nf & \yt \\
& JavaThread2 & \yt & \nf & \yt & \nf & \yt \\
& Looper1& \yt & \nf & \yt & \nf & \yt \\
\end{tabular}
\caption{DroidBench Results}
\label{tab:droidbench-res}
\end{table*}

\fi

Besides the quality of the results, also performances are important. Table~\ref{tab:performances} reports the mean and the median of the analysis times for the applications in DroidBench. As it turns out, \tool{} is one order of magnitude faster than both IccTA and AmanDroid, which in turn are one order of magnitude faster than DroidSafe. The extremely good performances of \tool{} are due to both design choices, like flow insensitivity on the activity life-cycle, and excellent support by Z3 in Horn clauses resolution. 

\begin{table}[htb]
\begin{center}
\begin{tabular}{c|c|c|c|c|}
\cline{2-5}
& \emph{IccTA} & \emph{AD} & \emph{DS} & \emph{HD} \\
\cline{1-5}
\multicolumn{1}{|c|}{\emph{Average Analysis Time}} & 19 & 11 & 176 & 1 \\ 
\cline{1-5}
\multicolumn{1}{|c|}{\emph{Median Analysis Time}} & 15 & 10 & 186 & 1 \\
\cline{1-5}
\end{tabular}
\end{center}
\caption{Analysis Time for DroidBench (Seconds)}
\label{tab:performances}
\end{table}

\subsection{Evaluation  on Real Applications}
In order to evaluate the practicality of our analysis, we performed a
test on the two largest applications available in the Google Play Top
30: the game Candy Crash Soda Saga (51.7 Mb) and the Facebook
application (46.5 Mb). We ran the experiments on a server with 64
multi-thread cores and 758 Gb of memory, however the highest memory
consumption by \tool{} was around 10 Gb, so it is possible to
reproduce our results even on a modern commercial machine. 

\tool{} found an information leak in Facebook, while Candy Crash Soda Saga
appears to be secure. The analysis took around 30 minutes and 60
minutes respectively. We tested all the existing competitors on both
applications, to check whether they could confirm the analysis
results. Unfortunately, AmanDroid crashed just after the beginning of
the analysis of Facebook, while both DroidSafe and IccTA failed to
terminate within the timeout we set (2 hours). We were able instead to
analyse Candy Crash Soda Saga using AmanDroid in around 50 minutes,
getting an information flow. After a manual inspection, we realized
this is a false positive due to the incorrect inclusion of the
\texttt{onHandleIntent} method of the class \texttt{IntentService}
among the possible sources of sensitive information: this is not
included in more recent proposals~\cite{GordonKPGNR15,LiEtAl15}. Both
IccTA and DroidSafe were not able to analyse the application within 2
hours. Due to space constraints, we refer to~\cite{full} for a more
comprehensive experimental evaluation on real applications. 

\subsection{Features and Limitations}
\label{sec:features}
As anticipated, the formalization in the previous sections only captures the \emph{core} of the analysis implemented in \tool{} and establishes the soundness of its principles. The tool, however, supports more features which are needed to make the analysis scale to real applications. We detail here some important aspects of \tool{} which are not covered by our formal model and we comment on current limitations.

\subsubsection*{Android Components}
Although the \sem{} model only represents activities and their life-cycle, \tool{} supports all the component types available on the Android platform, including services, broadcast receivers and content providers~\cite{Android}. The implementation of the analysis for these components does not significantly differ from the one for activities we presented in the paper, though it requires a correct modelling of their specific life-cycle.

\subsubsection*{Fragments}
Fragments are used to separate the functionality of an activity among different independent sub-components~\cite{Fragments}. In order to support a sound analysis of fragments, \tool{} over-approximates their life-cycle by executing all the fragments along with the containing activity in a flow-insensitive way. This might lead to precision problems on real applications, but this is the simplest of the sound options, which follows the philosophy we adopted for activity analysis.

\subsubsection*{Arrays} 
Though the static analysis we formalized is field-insensitive on
arrays, \tool{} supports a more precise treatment of array
indexes. Being value-sensitive, \tool{} statically approximates which
indexes of an array may be accessed at runtime: if a secret value is
stored in the first position of the array, but only the second element
of the array is leaked, the tool does not raise an alarm, contrarily
to all the other existing tools (cf. the breakdown on the experiments
in~\cite{full}).

\subsubsection*{Exceptions} 
\tool{} implements a conservative solution to handle exceptions, i.e., exceptions are always assumed to be thrown. A similar coarse over-approximation is implemented in FlowDroid~\cite{ArztRFBBKTOM14}. We leave a more precise treatment of exceptions to future work: we believe that the value-sensitivity of the analysis implemented in our tool will be crucial to limit the number of false alarms for exception handling. For instance, a value-sensitive analysis can ensure that a null pointer exception is never raised at runtime, since it over-approximates the set of the possible runtime values.

\subsubsection*{Inter-app Communication} 
\tool{} has limited support for inter-application communication, i.e., it conservatively detects an information leak whenever an intent storing secret data is sent to another application. More precise results could be achieved by analysing all the communicating applications simultaneously, but the current implementation of \tool{} only supports the analysis of a single application. We plan to leverage existing state-of-the-art solutions to overcome this limitation~\cite{LiEtAl15}. 

\subsubsection*{Threading}  
\tool{} handles multithreading  by assuming that threads are executed in a sequential,
but arbitrary order, much in the same spirit of the callbacks defining
the activity life-cycle. This is the same strategy used in FlowDroid.  We conjecture, but did not prove yet, that
this strategy is sound in our case, since the analysis is flow
insensitive on everything except for registers, which are not shared. For flow-sensitive analysis techniques (e.g.,
FlowDroid), instead, this strategy is in general unsound, since  it
may miss potential interleavings arising due to synchronization on
shared memory  (e.g., static heaps). The only aspect that should be added to our
static analysis is a thread pool simulation. In Java, every time the
method \texttt{execute} is called on a thread, this is placed in a
pool and then executed by the system by calling the runnable method
\texttt{run}. Our static analysis similarly binds each invocation of
\texttt{execute} to a corresponding \texttt{run} method.  

\subsubsection*{Reflection} 
Though supporting reflection soundly is an open research problem~\cite{SmaragdakisKB14}, \tool{} still covers a significant fraction of common reflection cases by implementing a simple string analysis. The solution we propose is in the same spirit of DroidSafe, i.e., reflective calls which can be statically resolved are replaced by direct calls to the appropriate method. Pragmatically, however, we observed that we are able to achieve much better results than DroidSafe for the reflection cases in DroidBench.

\subsubsection*{Limitations}
A comprehensive implementation of analysis stubs for method calls to the Android APIs is still lacking: we only implemented some selected stubs for our experiments, to show that our approach is feasible and practical. When a stub to an external library is missing, the tool tries to be conservative: the return value of the call is over-approximated to the top element of the corresponding abstract domain, and it is tainted whenever at least one of the arguments is tainted. 
Other important limitations of \tool{} are shared with existing
solutions~\cite{ArztRFBBKTOM14,GordonKPGNR15}. First, the analysis
does not capture  \emph{implicit} information flows at
present. Second, the analysis does not consider \emph{native
code}: this is a point we leave as a future work, observing that SMT solving has 
been successfully applied in the past to C code (see, e.g., the SLAM 
project~\cite{Ball_adecade}). Third, the analysis is oblivious to the \emph{semantics}
of the information flows, i.e., it lacks any built-in declassification mechanism
to qualify legitimate data flows. Since our analysis approximates data
information rather than just tracking taints, however, it is in principle 
possible to encode expressive data-dependent declassification policies, e.g.,
one could define the result of an encryption as untainted
only if the encryption is performed with the right key.


\section{Additional Related Work}
\label{sec:related}
Several papers have proposed an operational semantics for Android applications by now. The first attempt is due to Chaudhuri~\cite{Chaudhuri09}, who presented a core calculus to model Android applications. Later research proposed much more concrete models: Jeon \etal{} developed $\mu$-Dalvik, a relatively simple formal language which thoroughly models a significant fraction of the Dalvik opcodes~\cite{JeonMF12}. Wognsen \etal{} presented an even richer language, which also formalises exceptions and some common uses of reflection~\cite{WognsenKOH14}. Recently, Payet and Spoto complemented existing research by defining the first operational semantics for Android activities~\cite{PayetS14}. The semantics takes into account the event-driven behaviour of the activity lifecycle and, to some extent, the inter-component communication mechanism. Unfortunately, though, it represents only a small subset of the opcodes available in Dalvik and just models the control flow of activities, rather than the data flows enabled by inter-component communication. Our proposal integrates~\cite{JeonMF12} and~\cite{PayetS14}, while providing the first accurate description of how data flows between different components of an Android application.

Cassandra~\cite{LortzMSBSW14} is, to the best of our knowledge, the only tool 
implementing a provably sound information flow analysis for Android 
applications. The analysis is based on security types: well-typed programs ensure 
a termination-insensitive notion of non-interference, which proves the absence 
of both explicit and implicit information flows. By capturing implicit flows, 
Cassandra provides stronger security assurances than other static analysis tools, 
including ours. On the other hand, the analysis implemented in Cassandra is 
exclusively focused on the bytecode, and it does not track information leaks 
enabled by the application lifecycle. Moreover, the design of Cassandra is not 
very practical, since it requires application developers to write security 
certificates, giving a typing of all fields and methods in the application. Being 
type-based, Cassandra does not track any static approximation of runtime 
values, thus making it easy for malicious developers to force an overwhelming 
number of false alarms. We are not aware of any experimental evaluation of 
Cassandra so far.

Static analyses for improving the security of Android applications are not limited to information flow control: important applications include the detection of over-privileged apps~\cite{FeltCHSW11} and of attack surfaces for privilege escalation~\cite{BugliesiCS13}. Finally, it is worth mentioning that also dynamic analysis of Android applications is a popular research line~\cite{EnckGHTCCJMS14,JiaAFBSFKM13,TrippR14,HornyackHJSW11}. Dynamic analysis is largely complementary to static analysis, since it is typically more precise, but it hardly provides full coverage of all the possible execution paths and thus is not suitable to be employed in the vetting phase of an application.


\section{Conclusion}
\label{sec:conclusion}
We presented \tool, a tool for the static analysis of Android applications
based on Horn clause resolution. \tool is the first static analysis
tool for Android that comes with a formal proof of soundness covering
a large fragment of the Android ecosystem. Based on an available benchmark
proposed by the community, we experimentally showed that \tool is much more
efficient than competitors, very precise, and it is the first tool to 
detect all the existing (explicit) information flows.

Our approach  makes it easy to fine-tune the static
analysis, since one has just to modify the Horn clause generation
algorithm, while the resolution can be performed using off-the-shelf
SMT solvers, thus leveraging the tremendous progress in this field. 
In order to facilitate  future  extensions by the
community, we make our tool freely available, as source code as well
as through a web interface~\cite{full}.

We are currently working on the verification of CTL formulas, by 
using a recently developed encoding into Horn clauses~\cite{Beyene:2014}. Furthermore, we
plan to extend our tool in order to check non-interference properties and prove
the absence of implicit information flows. We would also like
to identify sound solutions to implement flow-sensitivity for heap
locations, thus making our static analysis even more precise. For
further boosting the precision, we intend to integrate in \tool a
recently developed string analysis engine for Z3~\cite{Trinh:2014}. 
Finally, we intend to extend the formal model and the proof of soundness in order
to cover the entire analysis.


\bibliographystyle{IEEEtranS}
\bibliography{local}

\iffull

\clearpage
\onecolumn
\appendix

\section{Formal Semantics of Statements}
\label{sec:bytecode}
We present an \emph{instrumented} semantics, which is useful for our soundness proof. With respect to the informal presentation in Section~\ref{sec:activity}, we need to extend the syntax of semantic domains as follows:
\[
\begin{array}{lcl}
L & \define & \locstate{\pp}{\stm^*}{R}{v^*} \\
\Sigma & \define & \methconf{\callstack}{\pi}{\heap}{\sheap}{\ell} 
\end{array}
\]
In the instrumented semantics, local states $L$ additionally contain a sequence of values $v^*$, representing the actual arguments provided upon method invocation when the local state was pushed on the call stack. Local configurations $\Sigma$, instead, are extended with a pointer $\ell$ to the activity modelled by the configuration.

\begin{definition}
Given a heap $\heap$, we let the partial function $\gettype{\heap}{v}$ be defined as follows:
\[
\gettype{\heap}{v} =
\begin{cases}
c & \text{if } v = \ell \wedge H(\ell) = \obj{c}{(f \mapsto v)^*} \\
\arrtype{\tau} & \text{if } v = \ell \wedge H(\ell) = \arr{\tau}{v^*} \\
\texttt{Intent} & \text{if } v = \ell \wedge H(\ell) = \intent{c}{(k \mapsto v)^*} \\
\primtype & \text{if } v = \prim 
\end{cases}
\]
where $\primtype$ is the type of the primitive value $\prim$.
\end{definition}

Let now $\super(c) = c'$ iff there exists $\class_i$ s.t. $\class_i = \cls{c}{c'}{c^*}{\field^*}{\method^*}$. Similarly, let $\interfaces(c) = \{c^*\}$ iff there exists $\class_i$ s.t. $\class_i = \cls{c}{c'}{c^*}{\field^*}{\method^*}$. Table~\ref{tab:subtyping} gives the subtyping rules for \sem, which are used, e.g., when defining the outcome of a type cast statement. Notice that array subtyping is covariant, which is unsound in presence of side-effects: like Java and the original presentation of $\mu$-Dalvik, we detect possible type errors at runtime.

\begin{table}[htb]
\begin{mathpar}
\inferrule[(Sub-Refl)]
{}
{\tau \subtype \tau}

\inferrule[(Sub-Trans)]
{\tau \subtype \tau' \\ \tau' \subtype \tau''}
{\tau \subtype \tau''}

\inferrule[(Sub-Ext)]
{}
{c \subtype \super(c)}

\inferrule[(Sub-Impl)]
{c' \in \interfaces(c)}
{c \subtype c'}

\inferrule[(Sub-Array)]
{\tau \subtype \tau'}
{\arrtype{\tau} \subtype \arrtype{\tau'}}
\end{mathpar}
\caption{Subtyping ($\tau \subtype \tau'$)}
\label{tab:subtyping}
\end{table}

Let $a[i] = v_i$ whenever $a = \arr{\tau}{v^*}$ and $o.f = v$ whenever $o = \obj{c}{(f_i \mapsto v_i)^*,f \mapsto v}$. Table~\ref{tab:rhs-eval} defines a convenience relation used to evaluate the right-hand side of a move instruction under a local configuration $\Sigma$. Notice that the evaluation of registers depends only on the top-most local state of the call stack of $\Sigma$.

\begin{table}[htb]
\begin{mathpar}
\inferrule*[lab=(Rhs-Register)]
{}
{\regval{r} = R(r)}

\inferrule*[lab=(Rhs-Array),width=10em]
{\ell = \regval{r_a} \\ a = \heap(\ell) \\ 
j = \regval{r_{\ind}}}
{\regval{r_a[r_{\ind}]} = a[j]}

\inferrule*[lab=(Rhs-Object),width=10em]
{\ell = \regval{r_o} \\ o = \heap(\ell)}
{\regval{r_o.f} = o.f}

\inferrule*[lab=(Rhs-Static)]
{}
{\regval{c.f} = \sheap(c.f)}

\inferrule*[lab=(Rhs-Prim)]
{}
{\regval{\prim} = \prim}
\end{mathpar}
\textbf{Convention:} in all the rules, let $\Sigma = \methconf{\callstack}{\pi}{\heap}{\sheap}{\_}$ with $\callstack = \locstate{\pp}{\stm^*}{R}{\_} :: \callstack'$.
\caption{Evaluation of Right-hand Sides ($\regval{\rhs} = v$)}
\label{tab:rhs-eval}
\end{table}

It is also useful to define substitutions for different syntactic categories, e.g., we let $o[f \mapsto v] = \obj{c}{(f_i \mapsto v_i)^*[f \mapsto v]}$ when $o = \obj{c}{(f_i \mapsto v_i)^*}$, and $\Sigma[\heap \mapsto \heap'] = \methconf{\callstack}{\pi}{\heap'}{\sheap}{\ell}$ when $\Sigma = \methconf{\callstack}{\pi}{\heap}{\sheap}{\ell}$. We do not provide full formal definitions for these substitutions, since their meaning will be clear from the context: it is only worth noticing that substitutions operating on elements of a local state only affect the \emph{top-most} local state of a local configuration $\Sigma$ when applied to it. For instance, given $\Sigma = \methconf{\callstack}{\pi}{\heap}{\sheap}{\ell}$ with $\callstack = \locstate{\pp}{\stm^*}{R}{v^*} :: \callstack'$, we let $\Sigma[R \mapsto R'] = \methconf{\callstack''}{\pi}{\heap}{\sheap}{\ell}$ where $\callstack'' = \locstate{\pp}{\stm^*}{R'}{v^*} :: \callstack'$, i.e., $\callstack'$ is unchanged.

We are finally ready to define the formal semantics of statements. Let $\Sigma = \methconf{\callstack}{\pi}{\heap}{\sheap}{\ell}$, we let $\getst{\Sigma} = \stm_{\pc}$ when $\callstack = \locstate{c,m,\pc}{\stm^*}{R}{\_} :: \callstack'$; we then let $\Sigma \rightsquigarrow \Sigma'$ if $\getst{\Sigma} = \stm$ and $\Sigma,\stm \Downarrow \Sigma'$ can be proved using the rules in Tables~\ref{tab:semantics}. There are only three perhaps surprising points: (1) when storing a value in an array cell, a dynamic check on the type of the value is performed, so as to ensure type soundness even in presence of the unsound subtyping rule for arrays; (2) when a new object is created, the pointer to it is annotated with the program point where creation takes place; and (3) upon method invocation, the value of the actual arguments is tracked in the syntax of the new local state. While (1) is an important aspect of the operational semantics, both (2) and (3) only serve static analysis purposes. Notice that we also use $\lookup$ to retrieve method bodies upon static calls: in this case, we assume $c' = c$.

\begin{table*}[p]\small
\begin{mathpar}
\inferrule*[lab=(R-Goto)]
{}
{\Sigma, \goto{\pc'} \Downarrow \Sigma[\pc \mapsto \pc']} 

\inferrule*[lab=(R-True)]
{\regval{r_1} \comp \regval{r_2}}
{\Sigma, \ifbr{r_1}{r_2}{\pc'} \Downarrow \Sigma[\pc \mapsto \pc']}

\inferrule*[lab=(R-False)]
{\neg(\regval{r_1} \comp \regval{r_2})}
{\Sigma, \ifbr{r_1}{r_2}{\pc'} \Downarrow \Sigma^+}

\inferrule*[width=10em,lab=(R-MoveReg)]
{v = \regval{\rhs} \\ R' = R[r \mapsto v]}
{\Sigma, \move{r}{\rhs} \Downarrow \Sigma^+[R \mapsto R']} 

\inferrule*[width=20em,lab=(R-MoveFld)]
{v = \regval{\rhs} \\ 
\ell = \regval{r_o} \\
o = \heap(\ell) \\ 
\heap' = \heap[\ell \mapsto o[f \mapsto v]]}
{\Sigma, \move{r_o.f}{\rhs} \Downarrow \Sigma^+[\heap \mapsto \heap']} 

\inferrule*[width=25em,lab=(R-MoveArr)]
{v = \regval{\rhs} \\ \ell = \regval{r_a} \\
\gettype{\heap}{\ell} = \arrtype{\tau} \\
\gettype{\heap}{v} \subtype \tau \\ 
a =  \heap(\ell) \\ j = \regval{r_\ind} \\
\heap' = \heap[\ell \mapsto a[j \mapsto v]] }
{\Sigma, \move{r_a[r_\ind]}{\rhs} \Downarrow \Sigma^+[\heap \mapsto \heap']}

\inferrule*[width=10em,lab=(R-MoveSFld)]
{v = \regval{\rhs} \\ \sheap' = \sheap[c'.f \mapsto v]}
{\Sigma, \move{c'.f}{\rhs} \Downarrow \Sigma^+[\sheap \mapsto \sheap']} 

\inferrule*[width=10em,lab=(R-UnOp)]
{v = \odot \regval{r_s} \\ R' = [r_d \mapsto v]}
{\Sigma, \unop{r_d}{r_s} \Downarrow \Sigma^+[R \mapsto R']}

\inferrule*[width=10em,lab=(R-BinOp)]
{v = \regval{r_1} \oplus \regval{r_2} \\ R' = R[r_d \mapsto v] }
{\Sigma, \binop{r_d}{r_1}{r_2} \Downarrow \Sigma^+[R \mapsto R']} 

\inferrule*[width=20em,lab=(R-NewObj)]
{o =  \obj{c'}{(f_{\tau} \mapsto \defvalue_{\tau})^*} \\ 
\ell = \pointer{p}{c,m,\pc} \notin \dom(\heap) \\
\heap' = \heap[\ell \mapsto o] \\ R' = R[r_d \mapsto \ell] }
{\Sigma, \new{r_d}{c'} \Downarrow \Sigma^+[\heap \mapsto \heap', R \mapsto R']}

\inferrule*[width=25em,lab=(R-NewArr)]
{\size = \regval{r_l} \\ 
a = \arr{\tau}{(\defvalue_{\tau})^{j \leq \size}} \\
\ell = \pointer{p}{c,m,\pc} \notin \dom(\heap) \\ 
\heap' = \heap[\ell \mapsto a] \\ 
R' = R[r_d \mapsto \ell]}
{\Sigma, \newarray{r_d}{r_l}{\tau}  \Downarrow \Sigma^+[\heap \mapsto \heap', R \mapsto R']}

\inferrule*[width=10em,lab=(R-Cast)]
{\ell = \regval{r_s} \\ \gettype{\heap}{\ell} \subtype \tau}
{\Sigma, \checkcast{r_s}{\tau} \Downarrow \Sigma^+}

\inferrule*[width=10em,lab=(R-InstOfTrue)]
{\ell = \regval{r_s} \\ \gettype{\heap}{\ell} \subtype \tau \\
R' = R[r_d \mapsto \true]}
{\Sigma, \instanceof{r_d}{r_s}{\tau} \Downarrow \Sigma^+[R \mapsto R']}

\inferrule*[width=10em,lab=(R-InstOfFalse)]
{\ell = \regval{r_s} \\ \gettype{\heap}{\ell} \not\subtype \tau \\
R' = R[r_d \mapsto \false]}
{\Sigma, \instanceof{r_d}{r_s}{\tau} \Downarrow \Sigma^+[R \mapsto R']}

\inferrule*[width=10em,lab=(R-Return)]
{\callstack = \locstate{c,m,\pc}{\_}{R}{\_} :: \locstate{\pp'}{\stm^*}{R'}{v^*} :: \callstack' \\
\callstack'' = \locstate{\pp'}{\stm^*}{R'[r_{\ret} \mapsto \regval{r_{\ret}}]}{v^*} :: \callstack'}
{\Sigma, \return \Downarrow \Sigma[\callstack \mapsto \callstack'']}

\inferrule*[width=30em,lab=(R-SCall)] 
{\lookup(c',m') = (c',\stm^*) \\ 
\sign(c',m') = \methsign{\tau_1,\ldots,\tau_n}{\tau}{\loc} \\
R' = ((r_j \mapsto \defvalue)^{j \leq \loc}, (r_{\loc+k} \mapsto \regval{r_k'})^{k \leq n}) \\ 
\callstack'' = \locstate{c',m',0}{\stm^*}{R'}{(\regval{r_k'})^{k \leq n}} :: \callstack^+}
{\Sigma, \sinvoke{c'}{m'}{r_1',\ldots,r_n'} \Downarrow \Sigma[\callstack \mapsto \callstack'']}

\inferrule*[width=30em,lab=(R-Call)]
{\ell = \regval{r_o} \\ \lookup(\gettype{\heap}{\ell},m') = (c',\stm^*) \\
\sign(c',m') = \methsign{\tau_1,\ldots,\tau_n}{\tau}{\loc} \\
R' = ((r_j \mapsto \defvalue)^{j \leq \loc}, r_{\loc + 1} \mapsto \ell, (r_{\loc+1+k} \mapsto 
\regval{r_k'})^{k \leq n}) \\ 
\callstack'' = \locstate{c',m',0}{\stm^*}{R'}{(\regval{r_k'})^{k \leq n}} :: \callstack^+}
{\Sigma, \invoke{r_o}{m'}{r_1',\ldots,r_n'} \Downarrow \Sigma[\callstack \mapsto \callstack'']}

\inferrule*[width=20em,lab=(R-NewIntent)]
{i =  \intent{c'}{\cdot} \\ 
\ell = \pointer{p}{c,m,\pc} \notin \dom(\heap) \\
\heap' = \heap[\ell \mapsto i] \\ R' = R[r_d \mapsto \ell]}
{\Sigma, \newintent{r_d}{c'} \Downarrow \Sigma^+[\heap \mapsto \heap', R \mapsto R']}

\inferrule*[width=20em,lab=(R-PutExtra)]
{\ell = \regval{r_i} \\ i = \heap(\ell) \\
k = \regval{r_k} \\ v = \regval{r_v} \\
\heap' = \heap[\ell \mapsto i[k \mapsto v]] }
{\Sigma, \putextra{r_i}{r_k}{r_v} \Downarrow \Sigma^+[\heap \mapsto \heap']}

\inferrule*[width=20em,lab=(R-GetExtra)]
{\ell = \regval{r_i} \\ k = \regval{r_k} \\
\heap(\ell) = i \\
\gettype{\heap}{i.k} \subtype \tau \\
v = i.k \\
R' = R[r_{\ret} \mapsto v]}
{\Sigma, \getextra{r_i}{r_k}{\tau} \Downarrow \Sigma^+[R \mapsto R']}

\inferrule*[width=25em,lab=(R-StartAct)]
{\ell = \regval{r_i} \\ H(\ell) = i \\
\pi' = i :: \pi }
{\Sigma, \startact{r_i} \Downarrow \Sigma^+[\pi \mapsto \pi']}
\end{mathpar}

\textbf{Convention:} in all the rules, let $\Sigma = \methconf{\callstack}{\pi}{\heap}{\sheap}{\_}$ with $\callstack = \locstate{c,m,\pc}{\_}{R}{\_} :: \callstack_0$. We let $\Sigma^+$ (resp. $\alpha^+$) stand for $\Sigma$ (resp. $\alpha$) where $\pc$ is replaced by $\pc + 1$.
\caption{Concrete small step semantics of \sem{} ($\Sigma,\stm \Downarrow \Sigma'$) - Statements}
\label{tab:semantics}
\end{table*}


\clearpage

\section{Taint Analysis Specification}
\label{sec:taint}
First, we extend the semantic domains specification from
Table~\ref{tab:dalvik-domains} and the corresponding abstract domains
specification from Table~\ref{tab:absdoms}:
\[
\begin{array}{lccclcc}
\text{Taints} & \taint  \define  \secret ~|~ \public & & & & &\\
\text{Values} & u,v \define  \prim^\taint ~|~ \ell & & \text{Abs. values} & \absual,\absval & \define & \emptyset ~|~ \{\absprim^\taint\} ~|~ \{\absloc\} ~|~ \absval \cup \absval 
\end{array}
\]
We introduce a definition of taint $\taint$ that ranges over
$\public$ and $\secret$ values forming two-valued lattice with $\secret$ as
the top element. Plus, primitive values $\prim$ and their abstractions are extended with a taint
annotation.

Second, we define a taint function $\taintf$ that extracts taints from
values and its abstract version $\ataintf$ as
follows:
\[
\begin{array}{ll}
\taintf(v) =
\begin{cases}
& \text{if } v = \ell \wedge H(\ell) = \obj{c}{(f \mapsto v_i)^*} \\
\bigsqcup_i \taintf(v_i) & \text{if } v = \ell \wedge H(\ell) = \arr{\tau}{v_i^*} \\
& \text{if } v = \ell \wedge H(\ell) = \intent{c}{(k \mapsto v_i)^*} \\
\taint & \text{if } v = \prim^\taint 
\end{cases}
&\ataintf(\absval) =
\begin{cases}
\bigsqcup_i \ataintf(\absval_i) & \text{if } \absval = \{ \absloc \} \wedge\absheap(\absloc, \absobj{c}{(f \mapsto \absval_i)^*}) \\
\ataintf(\absval_i) & \text{if } \absval = \{ \absloc \} \wedge \absheap(\absloc, \absarray{\tau}{\absval_i})\\
\ataintf(\absval_i) & \text{if } \absval = \{ \absloc \} \wedge
\absheap(\absloc, \absintent{c}{\absval_i})\\
\taint & \text{if } \absval = \{ \absprim^\taint \}  \\
\ataintf(\absval_i) \sqcup \ataintf(\absval_j) & \text{if } \absval =
\absval_i \cup \absval_j \\
\public & \text{if } \absval = \emptyset
\end{cases}
\end{array}
\]
The taint function $\taintf$ returns taint annotations for
primitive values. In case of a pointer it returns a join of
taints that can be accessed by the pointer. The abstract taint
function $\ataintf$ is defined in a similar way, with extensions for
the empty set, which is $\public$ and the union of two abstract
values, for which we have again the join of the taints.

The specification of the taint propagation logic for the value amounts to changing the binary
and unary operations for both concrete and abstract semantics as follows:
\[
\begin{array}{l|cc}
v_d = v_1 \oplus v_2 &\taintf(v_d) = \taintf(v_1) \sqcup
                        \taintf(v_2) & \ataintf(\absval_d) = \ataintf(\absval_1) \sqcup
                        \ataintf(\absval_2)\\
v_d = \odot v_s &\taintf(v_d) = \taintf(v_s) & \ataintf(\absval_d) =
                                          \ataintf(\absval_s)
\end{array}
\]
In case of a binary operation the taint value of the destination
register is raised to the highest taint among the values of the registers
used in the operation. The result of a unary operation has the same taint
as the source.

Also we assume to have two sets of pairs \sinks{} and
\sources, that contain a pair ($c$, $m$), if a method $m$ of a class $c$
is a sink/source respectively We assume that when a source return a value, it always
has $\secret$ taint.

\begin{definition}
\label{def:leak}
Let $\Psi$ be the initial configuration of a program $P$, we say that
$P$ \emph{leaks} starting from $\Psi$ if and only if there exists $(c,m) \in \sinks$ such
that $\Psi \Rightarrow^* \actconf{\actstack}{\heap}{\sheap}$ and
$\actstack$ contains an active frame
$\uactframe{\ell}{s}{\pi}{\callstack}$ such that $\callstack =
\locstate{c,m,0}{\stm^*}{R}{\_} :: \callstack'$, $R(r_k) =
v$ and $\taintf(v) = \secret$ for some $r_k$ and some $v$. 
\end{definition}

\begin{lemma}
\label{lem:no-leak}
If for all sinks $(c, m) \in \sinks: \translate{P} \cup \rfconf(\Psi)
\vdash \absreg{c, m, 0}{\_}{\absval^*}$ we have $\ataintf(\absval_i)
= \public$ for each $i$, then program $P$ does not leak
starting from $\Psi$.
\end{lemma}
\begin{IEEEproof}
We prove the contrapositive. Assume that a program $P$ satisfies Definition~\ref{def:leak}, then there exists a
configuration $\Psi'$,
starting from $\Psi$, where one of the registers $r_k$ in a sink $(c,m)$ 
contains a $\secret$ value. By Theorem~\ref{trm:preservation} there exists
$\absprog :> \rfconf(\Psi')$ such that $\translate{P} \cup
\rfconf(\Psi) \vdash \absprog$. The relation $\absprog :> \rfconf(\Psi')$
can only hold if $\absreg{c, m, 0}{\_}{\absval^*} \in \absprog$ and $\ataintf(\absval_k)
= \secret$. 
\end{IEEEproof}


\clearpage

\section{Soundness Proofs}
\label{sec:proofs}
\subsection{Representation Functions}
We presuppose the existence of a representation function $\rfprim$ which associates to each primitive value $\prim$ a corresponding abstract value $\{\absprim\}$. For a location $\ell = \pointer{p}{\ann}$, we let $\rfloc(\ell) = \{\ann\}$. Based on this, we define $\rfval(v)$ as follows:
\[
\rfval(v) =
\begin{cases}
\rfprim(v) & \text{if } v = \prim \\
\rfloc(v) & \text{if } v = \ell
\end{cases}
\]
We typically omit brackets around singleton abstract values. We then define $\rfblock(b)$ as follows:
\[
\rfblock(b) =
\begin{cases}
\absobj{c}{(f \mapsto \absval)^*} & \text{if } b =  \obj{c}{(f \mapsto v)^*} \text{ and } \forall i: \rfval(v_i) = \absval_i \\
\absintent{c}{\absval} & \text{if } b = \intent{c}{(f \mapsto v)^*} \text{ and } \absval = \sqcup_i\, \rfval(v_i) \\
\absarray{\tau}{\absval} & \text{if } b = \arr{\tau}{v^*} \text{ and } \absval = \sqcup_i\, \rfval(v_i)
\end{cases}
\]
Using these definitions, we can define how configurations are translated into facts by a corresponding representation function. This requires one to define a number of clauses, summarized below:

\[
\begin{array}{lcl}
\rflocstate(\locstate{c,m,\pc}{\stm^*}{R}{u^*}) & = & \{\absreg{c,m,\pc}{\absual^*}{\absval^*} ~|~ \forall j: \absual_j = \rfval(u_j) \wedge \forall k: \absval_k = \rfval(R(r_k))\} \cup \bigcup_i\, \ainstfull{\stm_i}{c,m,i} \\

\rfcall(\callstack) & = & \bigcup_{i \in [1,n]} \rflocstate(L_i) \text{ whenever } \callstack = L_1 :: \ldots :: L_n \\

\rfheap(\heap) & = & \{\absheap(\absloc,\absblock) ~|~ \heap = \heap',\ell \mapsto b \wedge \absloc = \rfloc(\ell) \wedge \absblock = \rfblock(b)\} \\

\rfstat(\sheap) & = & \{\abssheap(c,f,\absval) ~|~ \sheap = \sheap',c.f \mapsto v \wedge \absval = \rfval(v)\} \\

\rfdispatch{\ell}(\pi) & = & \{\absdispatch(c,\absblock) ~|~ c = \rfloc(\ell) \wedge \pi = \pi_0 :: i :: \pi_1 \wedge \absblock = \rfblock(i)\} \\

\rflconf(\methconf{\callstack}{\pi}{\heap}{\sheap}{\ell}) & = & \rfcall(\callstack) \cup \rfdispatch{\ell}(\pi) \cup \rfheap(\heap) \cup \rfstat(\sheap) \\

\rfframe(\actframe{\ell}{s}{\pi}{\callstack}) & = & \rfframe(\uactframe{\ell}{s}{\pi}{\callstack}) = \rfdispatch{\ell}(\pi) \cup \rfcall(\callstack) \\

\rfastk(\actstack) & = & \bigcup_{i \in [1,n]} \rfframe(\varphi_i) \text{ whenever } \actstack = \varphi_1 :: \ldots :: \varphi_n \\

\rfconf(\actconf{\actstack}{\heap}{\sheap}) & = & \rfastk(\actstack) \cup \rfheap(\heap) \cup \rfstat(\sheap)
\end{array}
\]

\subsection{Ordering Abstract Values and Facts}
We presuppose the existence of a pre-order $\poprim$ on primitive singleton abstract values. Based on this, we define a pre-order $\poval$ on abstract values by having $\absual \poval \absval$ iff:
\begin{itemize}
\item $\forall \absprim \in \absual: \exists \absprim' \in \absval: \absprim \poprim \absprim'$;
\item $\forall \absloc \in \absual: \absloc \in \absval$.
\end{itemize}
We then build a pre-order $\poseq$ on sequences of abstract values by having $\absual^* \poseq \absval^*$ iff $\absual^*$ and $\absval^*$ have the same length and:
\[
\forall i: \absual_i \poval \absval_i.
\]
We can then define a pre-order $\poblk$ on abstract blocks as follows:
\begin{itemize}
\item if $\absblock = \absobj{c}{(f \mapsto \absual)^*}$ and $\absblock' = \absobj{c}{(f \mapsto \absval)^*}$ and $\absual^* \poseq \absval^*$, then $\absblock \poblk \absblock'$;
\item if $\absblock = \absintent{c}{\absual}$ and $\absblock' = \absintent{c}{\absval}$ and $\absual \poval \absval$, then $\absblock \poblk \absblock'$;
\item if $\absblock = \absarray{\tau}{\absual}$ and $\absblock' = \absarray{\tau} {\absval}$ and $\absual \poval \absval$, then $\absblock \poblk \absblock'$.
\end{itemize}

Finally, we let $\fact \sqsubseteq \fact'$ be the least pre-order on facts such that:
\begin{itemize}
\item $\absreg{c,m,\pc}{\absual_{call}^*}{\absual^*} \sqsubseteq \absreg{c,m,\pc}{\absval_{call}^*}{\absval^*}$ whenever $\absual_{call}^* \poseq \absval_{call}^*$ and $\absual^* \poseq \absval^*$;
\item $\absheap(\absloc,\absblock) \sqsubseteq \absheap(\absloc,\absblock')$ whenever $\absblock \poblk \absblock'$;
\item $\abssheap(c,f,\absual) \sqsubseteq \abssheap(c,f,\absval)$ whenever $\absual \poval \absval$;
\item $\prhs{\absual} \sqsubseteq \prhs{\absval}$ whenever $\absual \poval \absval$;
\item $\absresult{c,m}{\absual_{call}^*}{\absual^*} \sqsubseteq \absresult{c,m}{\absval_{call}^*}{\absval^*}$ whenever $\absual_{call}^* \poseq \absval_{call}^*$ and $\absual^* \poseq \absval^*$;
\item $\absdispatch(c,\absblock) \sqsubseteq \absdispatch(c,\absblock')$ whenever $\absblock \poblk \absblock'$.
\end{itemize}

\subsection{Formal Results}

\subsubsection{Preliminaries}

\begin{definition}
A local configuration $\Sigma = \methconf{\callstack}{\pi}{\heap}{\sheap}{\ell}$ is \emph{well-formed} if and only if, whenever $\callstack = L_1 :: \ldots :: L_n$, we have:
\begin{itemize}
\item either $n \in \{0,1\}$, i.e., $\callstack$ is either empty or it contains just a single local state;
\item or $n \geq 2$ and for each $i \in [2,n]$, either of the
  following conditions hold true:
\MESSAGEIfor{MS}{151202}{We replace $\stm_{\pc-1}$ with $\stm_{\pc}$, without
  that analysis cannot over-approximate representation function (i.e.,
  we
  will have function calls treated as both nops and real calls in the
  representation function).}
\begin{itemize}
\item $L_i = \locstate{c',m',\pc'}{\stm'^*}{R'}{v^*}$ and $L_{i-1} = \locstate{c,m,\pc}{\stm^*}{R}{\_}$ with $\stm_{\pc} = \invoke{r_o}{m'}{r_1',\ldots,r_n'}$, \\ $\lookup(\gettype{\heap}{\regval{r_o}},m') = (c',\stm'^*)$, $\sign(c',m') = \methsign{\tau_1,\ldots,\tau_n}{\tau}{\loc}$ and $v^* = (\regval{r_k'})^{k \leq n}$
\item $L_i = \locstate{c',m',\pc'}{\stm'^*}{R'}{v^*}$ and $L_{i-1} = \locstate{c,m,\pc}{\stm^*}{R}{\_}$ with $\stm_{\pc} = \sinvoke{c'}{m'}{r_1',\ldots,r_n'}$, \\ $\lookup(c',m') = (c',\stm'^*)$, $\sign(c',m') = \methsign{\tau_1,\ldots,\tau_n}{\tau}{\loc}$ and $v^* = (\regval{r_k'})^{k \leq n}$.
\end{itemize}
\end{itemize} 
\end{definition}

\begin{lemma}[Preserving Local Well-formation]
\label{lem:preserve-local}
If $\Sigma$ is well-formed and $\Sigma \rightsquigarrow^* \Sigma'$, then $\Sigma'$ is well-formed.
\end{lemma}
\begin{IEEEproof}
By induction on the length of the reduction sequence and a case analysis on the last rule applied.
\end{IEEEproof}

\begin{definition}
A heap $\heap$ is \emph{well-typed} if and only if, whenever $\heap(\ell) = \obj{c}{(f_i \mapsto v_i)^{i \leq n}}$, for all $i \in [1,n]$ we have $\gettype{\heap}{v_i} \subtype \tau_i$, where $\tau_i$ is the declared type of field $f_i$ for an object of type $c$ according to the underlying program.
\end{definition}

\begin{assumption}[Java Type Soundness]
\label{asm:java-sound}
If $\methconf{\callstack}{\pi}{\heap}{\sheap}{\ell} \rightsquigarrow \methconf{\callstack'}{\pi'}{\heap'}{\sheap'}{\ell}$, then for any value $v$ we have $\gettype{\heap'}{v} \subtype \gettype{\heap}{v}$. Moreover, if $\heap$ is well-typed, then also $\heap'$ is well-typed.
\end{assumption}

\begin{definition}
A configuration $\Psi = \actconf{\actstack}{\heap}{\sheap}$ is \emph{well-formed} if and only if:
\begin{itemize}
\item whenever $\actstack = \actstack_0 :: \varphi :: \actstack_1$ with $\varphi \in \{\actframe{\ell}{s}{\pi}{\callstack},\uactframe{\ell}{s}{\pi}{\callstack}\}$, we have $\heap(\ell) = \obj{c}{(f \mapsto v)^*}$ for some activity class $c$ and $\ell = \pointer{p}{c}$ for some pointer $p$;

\item whenever $\actstack = \actstack_0 :: \varphi :: \actstack_1$ with $\varphi \in \{\actframe{\ell}{s}{\pi}{\callstack},\uactframe{\ell}{s}{\pi}{\callstack}\}$, we have that $\Sigma = \methconf{\callstack}{\pi}{\heap}{\sheap}{\ell}$ is a well-formed local configuration;

\item $\heap$ is a well-typed heap.
\end{itemize}
\end{definition}

\begin{lemma}[Preserving Well-formation]
\label{lem:preserve-well}
If $\Psi$ is well-formed and $\Psi \Rightarrow^* \Psi'$, then $\Psi'$ is well-formed.
\end{lemma}
\begin{IEEEproof}
By induction on the length of the reduction sequence and a case analysis on the last rule applied, using Lemma~\ref{lem:preserve-local} and Assumption~\ref{asm:java-sound} to deal with case \irule{A-Active}.
\end{IEEEproof}
From now on, we tacitly focus only on well-formed configurations. All the formal results only apply to them: notice that well-formed configurations always reduce to well-formed configurations by Lemma~\ref{lem:preserve-well}.

\subsubsection{Main Results}

\begin{lemma}
\label{lem:sub-order}
If $\absprog \subseteq \absprog'$, then $\absprog <: \absprog'$.
\end{lemma}

\begin{lemma}
\label{lem:trans-order}
If $\absprog <: \absprog'$ and $\absprog' <: \absprog''$, then $\absprog <: \absprog''$.
\end{lemma}

\begin{lemma}
\label{lem:join-order}
If $\absprog_1 <: \absprog_2$ and $\absprog_3 <: \absprog_4$, then $\absprog_1 \cup \absprog_3 <: \absprog_2 \cup \absprog_4$.
\end{lemma}

\begin{assumption}[Soundness of the Abstract Operations]
\label{asm:sound-op}
We assume all the following properties:
\begin{itemize}
\item if $u \comp v$, then $\absual\ \acomp\ \absval$ for any $\absual,\absval$ such that $\absual :> \rfval(u)$ and $\absval :> \rfval(v)$
\item for any $\absval :> \rfval(v)$, we have $\aunop \absval :> \rfval(\odot v)$
\item for any $\absual,\absval$ such that $\absual :> \rfval(u)$ and $\absval :> \rfval(v)$, we have $\absual\ \abinop\ \absval :> \rfval(u \oplus v)$
\end{itemize}
\end{assumption}

\begin{assumption}[Overriding]
\label{asm:overriding}
If $\lookup(c,m) = (c',\stm^*)$, then $c \subtype c'$.
\end{assumption}

In the next results, let $\absprog \vdash \absprog'$ whenever $\absprog \vdash \fact$ for each $\fact \in \absprog'$.

\begin{lemma}[Right-hand Sides]
\label{lem:rhs}
Let $\Sigma = \methconf{\callstack}{\pi}{\heap}{\sheap}{\ell}$ with $\callstack = \locstate{\pp}{\stm^*}{R}{u^*}$ and let $\regval{\rhs} = v$, then for any $\absprog :> \rflconf(\Sigma)$ there exists $\absval$ such that $\rfval(v) \poval \absval$ and $\absprog \cup \arhs{\rhs} \vdash \prhs{\absval}$.
\end{lemma}
\begin{IEEEproof}
By a case analysis on the structure of $\rhs$.
\end{IEEEproof}

\begin{lemma}[Local Preservation]
\label{lem:local}
If $\Sigma \rightsquigarrow \Sigma'$ under a given program $P$, then for any $\absprog :> \rflconf(\Sigma)$ there exists $\absprog' :> \rflconf(\Sigma')$ such that $\translate{P} \cup \absprog \vdash \absprog'$.
\end{lemma}
\begin{IEEEproof}
(Sketch) By a case analysis on the rule applied in the reduction step. The cases for the $\texttt{move}$ instruction use Lemma~\ref{lem:rhs}. The case for the $\return$ instruction exploits the (implicit) well-formation assumption of the local configuration $\Sigma$. The case for the $\texttt{invoke}$ instruction uses Assumption~\ref{asm:overriding}. The cases for comparison operators and primitive operations exploit Assumption~\ref{asm:sound-op}.
\end{IEEEproof}

\begin{lemma}[Serialization]
\label{lem:serialization}
Both the following statements hold true:
\begin{itemize}
\item if $\serval{\heap}(v) = (v',\heap')$, then $\rfval(v) = \rfval(v')$
\item if $\serblock{\heap}(b) = (b',\heap')$, then $\rfblock(b) = \rfblock(b')$
\end{itemize}
\end{lemma}
\begin{IEEEproof}
If $v = \prim$, then $v' = \prim$ and $\rfval(v) = \rfval(v') = \rfprim(\prim)$. If $v = \pointer{p}{\ann}$, then $v' = \pointer{p'}{\ann}$ for some pointer $p'$ and $\rfval(v) = \rfloc(\pointer{p}{\ann}) = \ann = \rfloc(\pointer{p'}{\ann}) = \rfval(v')$. The second point is a direct consequence of the first one.
\end{IEEEproof}

\MESSAGEIfor{MS}{151202}{New definition for the size function.}
\begin{definition}
We define a function $\bvsize{\heap}$ which assigns to values and blocks a natural number as follows:
\begin{itemize}
\item $\serialized \vdash \bvsize{\heap}(\prim) = 1$
\item $\ell \notin \serialized; \serialized,\ell \vdash
  \bvsize{\heap}(\ell) = 1 + \bvsize{\heap}(\heap(\ell))$
\item $\ell \in \serialized; \serialized,\ell \vdash \bvsize{\heap}(\ell) = 0$
\item $\serialized \vdash \bvsize{\heap}(\obj{c}{(f_i \mapsto v_i)^*}) = 1 + \sum_i \bvsize{\heap}(v_i)$
\item $\serialized \vdash \bvsize{\heap}(\intent{c}{(k_i \mapsto v_i)^*}) = 1 + \sum_i \bvsize{\heap}(v_i)$
\item $\serialized \vdash \bvsize{\heap}(\arr{\tau}{v^*}) = 1 + \sum_i \bvsize{\heap}(v_i)$
\end{itemize}
\end{definition}

\begin{lemma}[Heap Serialization]
\label{lem:heap-serialization}
If $\absprog :> \rfheap(\heap)$, then:
\begin{itemize}
\item $\serval{\heap}(v) = (v',\heap')$ implies $\absprog :> \rfheap(\heap')$
\item $\serblock{\heap}(b) = (b',\heap')$ implies $\absprog :> \rfheap(\heap')$
\end{itemize}
\end{lemma}
\begin{IEEEproof}
By simultaneous induction on the size of the syntactic element in the
antecedent. If $v = \prim$, then $\heap'$ is empty, hence
$\rfheap(\heap') = \emptyset$ and we are done. If $v =
\pointer{p}{\ann}$, then $\heap' = \heap'',\pointer{p'}{\ann} \mapsto
b$ with $\serblock{\heap}(\heap(\pointer{p}{\ann})) = (b, \heap'')$
and $v' = \pointer{p'}{\ann}$. By induction hypothesis $\absprog :> \rfheap(\heap'')$, so to conclude we just need to show that:
\[
\begin{array}{lcll}
\absprog & :> & \rfheap(\pointer{p'}{\ann} \mapsto b) \\
 & = & \{\absheap(\ann,\rfblock(b))\} & \text{by definition} \\
 & = & \{\absheap(\ann,\rfblock(\heap(\pointer{p}{\ann})))\} & \text{by Lemma~\ref{lem:serialization}} \\
 & = & \rfheap(\pointer{p}{\ann} \mapsto \heap(\pointer{p}{\ann})) & \text{by definition}
\end{array}
\]
but this follows from the hypothesis $\absprog :> \rfheap(\heap)$. The remaining cases for blocks follow by inductive hypothesis.
\end{IEEEproof}

\begin{theorem}[Preservation]
\label{trm:preservation}
If $\Psi \Rightarrow^* \Psi'$ under a given program $P$, then there exists $\absprog :> \rfconf(\Psi')$ such that $\translate{P} \cup \rfconf(\Psi) \vdash \absprog$.
\end{theorem}
\begin{IEEEproof}
By induction on the length of the reduction sequence. If the reduction sequence is empty, we have $\Psi' = \Psi$ and the result follows by picking $\absprog = \rfconf(\Psi)$. Otherwise, assume that $\Psi \Rightarrow^* \actconf{\actstack}{\heap}{\sheap}$ in $n \geq 0$ reduction steps and let $\actconf{\actstack}{\heap}{\sheap} \Rightarrow \actconf{\actstack'}{\heap'}{\sheap'}$. By induction hypothesis there exists $\absprog' :> \rfconf(\actconf{\actstack}{\heap}{\sheap})$ such that $\translate{P} \cup \rfconf(\Psi) \vdash \absprog'$, we show that there exists $\absprog$ such that $\absprog :> \rfconf(\actconf{\actstack'}{\heap'}{\sheap'})$ and $\translate{P} \cup \rfconf(\Psi) \vdash \absprog$. The proof is by a case analysis on the rule applied in the last reduction step:
\begin{itemize}
\item[\irule{A-Active}]: let $\actstack = \actstack_0 :: \uactframe{\ell}{s}{\pi}{\callstack} :: \actstack_1$ and $\actstack' = \actstack_0 :: \uactframe{\ell}{s}{\pi'}{\callstack'} :: \actstack_1$ with $\methconf{\callstack}{\pi}{\heap}{\sheap}{\ell} \rightsquigarrow \methconf{\callstack'}{\pi'}{\heap'}{\sheap'}{\ell}$. Since $\rflconf(\methconf{\callstack}{\pi}{\heap}{\sheap}{\ell}) \subseteq \rfconf(\actconf{\actstack}{\heap}{\sheap})$, we have $\rflconf(\methconf{\callstack}{\pi}{\heap}{\sheap}{\ell}) <: \rfconf(\actconf{\actstack}{\heap}{\sheap})$ by Lemma~\ref{lem:sub-order}. Since $\rflconf(\methconf{\callstack}{\pi}{\heap}{\sheap}{\ell}) <: \rfconf(\actconf{\actstack}{\heap}{\sheap})$ and $\rfconf(\actconf{\actstack}{\heap}{\sheap}) <: \absprog'$, we get $\rflconf(\methconf{\callstack}{\pi}{\heap}{\sheap}{\ell}) <: \absprog'$ by Lemma~\ref{lem:trans-order}. 
Hence, by Lemma~\ref{lem:local} there exists $\absprog'' :> \rflconf(\methconf{\callstack'}{\pi'}{\heap'}{\sheap'}{\ell})$ such that $\translate{P} \cup \absprog' \vdash \absprog''$. By the weakening property of the logic, the latter implies $\translate{P} \cup \rfconf(\Psi) \cup \absprog' \vdash \absprog''$. Since we have $\translate{P} \cup \rfconf(\Psi) \vdash \absprog'$ and $\translate{P} \cup \rfconf(\Psi) \cup \absprog' \vdash \absprog''$, we get $\translate{P} \cup \rfconf(\Psi) \vdash \absprog''$ by the admissibility of the cut rule. Recall now that $\absprog'' :> \rflconf(\methconf{\callstack'}{\pi'}{\heap'}{\sheap'}{\ell}) = \rfcall(\callstack') \cup \rfdispatch{\ell}(\pi') \cup \rfheap(\heap') \cup \rfstat(\sheap')$, so we have:
\begin{itemize}
\item[(1)] $\absprog'' :> \rfcall(\callstack')$
\item[(2)] $\absprog'' :> \rfdispatch{\ell}(\pi')$
\item[(3)] $\absprog'' :> \rfheap(\heap')$
\item[(4)] $\absprog'' :> \rfstat(\sheap')$
\end{itemize}
We then observe that $\translate{P} \cup \rfconf(\Psi) \vdash \absprog' :> \rfconf(\actconf{\actstack}{\heap}{\sheap})$, which similarly implies:
\begin{itemize}
\item[(5)] $\absprog' :> \rfastk(\actstack_0)$
\item[(6)] $\absprog' :> \rfastk(\actstack_1)$
\end{itemize}
Combining all these facts, we get $\absprog' \cup \absprog'' :> \rfconf(\actconf{\actstack'}{\heap'}{\sheap'})$ by Lemma~\ref{lem:join-order}. Given that $\translate{P} \cup \rfconf(\Psi) \vdash \absprog' \cup \absprog''$, we conclude the case;

\item[\irule{A-Deactivate}]: in this case $\rfconf(\actconf{\actstack}{\heap}{\sheap}) = \rfconf(\actconf{\actstack'}{\heap'}{\sheap'})$, hence the conclusion immediately follows by the induction hypothesis;

\item[\irule{A-Step}]: let $\actstack = \actframe{\ell}{s}{\pi}{\ocallstack} :: \actstack_0$ and $\actstack' = \uactframe{\ell}{s'}{\pi}{\getcb{\ell}{s'}} :: \actstack_0$ for some $(s,s') \in \lifecycle$, $\heap' = \heap$ and $\sheap' = \sheap$. Since $\translate{P} \cup \rfconf(\Psi) \vdash \absprog' :> \rfconf(\actconf{\actstack}{\heap}{\sheap})$, we have:
\begin{itemize}
\item[(1)] $\absprog' :> \rfastk(\actstack_0)$
\item[(2)] $\absprog' :> \rfdispatch{\ell}(\pi)$
\end{itemize}
Since we only focus on well-formed configurations, we have $\heap(\ell) = \obj{c}{(f \mapsto u)^*}$ for some activity class $c$ and $\ell = \pointer{p}{c}$ for some pointer $p$. We then observe that $\getcb{\ell}{s'} = \locstate{c',m,0}{\stm^*}{R}{v^*} :: \varepsilon$, where $(c',\stm^*) = \lookup(c,m)$ for some $m \in \cb(c,s)$, $\sign(c',m) = \methsign{\tau_1,\ldots,\tau_n}{\tau}{\loc}$ and:
\[ 
R = ((r_i \mapsto \defvalue)^{i \leq \loc}, r_{\loc+1} \mapsto \ell, (r_{\loc+1+j} \mapsto v_j)^{j \leq n}),
\]
for some values $v_1,\ldots,v_n$ of the correct type $\tau_1,\ldots,\tau_n$. By Assumption~\ref{asm:overriding}, we also have $c \subtype c'$.

Given that $\absprog' :> \rfconf(\actconf{\actstack}{\heap}{\sheap})$, we have $\absprog' :> \rfheap(\heap)$, which implies that there exists $\absheap(\absloc,\absblock) \in \absprog'$ such that $\absloc = \rfloc(\ell) = c$ and $\absblock \sqsupseteq \rfblock(\obj{c}{(f \mapsto u)^*})$. This implies that $\absblock = \absobj{c}{(f \mapsto \absval)^*}$ for some $v^*$ such that $\forall i: \absval_i \sqsupseteq \rfval(u_i)$. Since $\translate{P} \cup \rfconf(\Psi) \vdash \absprog'$ and $\absheap(\absloc,\absblock) = \absheap(c,\absobj{c}{(f \mapsto \absval)^*}) \in \absprog'$, we have in particular $\translate{P} \cup \rfconf(\Psi) \vdash \absheap(c,\absobj{c}{(f \mapsto \absval)^*})$, hence:
\[
\translate{P} \cup \rfconf(\Psi) \vdash \absreg{c',m,0}{(\top_{\tau_j})^{j \leq n}}{(\adefvalue)^{k \leq \loc},c,(\top_{\tau_j})^{j \leq n}},
\]
by using the implications $\rulename{Cbk}$ included in $\translate{P}$. We then observe that:
\[
\{\absreg{c',m,0}{(\top_{\tau_j})^{j \leq n}}{(\adefvalue)^{k \leq \loc},c,(\top_{\tau_j})^{j \leq n}}\} :> \rfcall(\getcb{\ell}{s'})
\]
By combining (1), (2) and the last observation through Lemma~\ref{lem:join-order} we then get:
\[
\{\absreg{c',m,0}{(\top_{\tau_j})^{j \leq n}}{(\adefvalue)^{k \leq \loc},c,(\top_{\tau_j})^{j \leq n}}\} \cup \absprog' :> \rfcall(\getcb{\ell}{s'}) \cup \rfastk(\actstack_0) \cup \rfdispatch{\ell}(\pi) = \rfastk(\actstack')
\]
Since $\translate{P} \cup \rfconf(\Psi) \vdash \{\absreg{c',m,0}{(\top_{\tau_j})^{j \leq n}}{(\adefvalue)^{k \leq \loc},c,(\top_{\tau_j})^{j \leq n}}\} \cup \absprog'$, we conclude the case;

\item[\irule{A-Destroy}]: in this case $\rfconf(\actconf{\actstack'}{\heap'}{\sheap'}) \subseteq \rfconf(\actconf{\actstack}{\heap}{\sheap})$, hence $\rfconf(\actconf{\actstack'}{\heap'}{\sheap'}) <: \rfconf(\actconf{\actstack}{\heap}{\sheap})$ by Lemma~\ref{lem:sub-order}. Since $\rfconf(\actconf{\actstack'}{\heap'}{\sheap'}) <: \rfconf(\actconf{\actstack}{\heap}{\sheap})$ and $\rfconf(\actconf{\actstack}{\heap}{\sheap}) <: \absprog'$, we have $\rfconf(\actconf{\actstack'}{\heap'}{\sheap'}) <: \absprog'$ by Lemma~\ref{lem:trans-order}. Given that $\translate{P} \cup \rfconf(\Psi) \vdash \absprog'$, we conclude the case;

\item[\irule{A-Back}]: let $\actstack' = \actstack = \actframe{\ell}{\actstate{running}}{\varepsilon}{\ocallstack} :: \actstack_0$, $\heap' = \heap[\ell \mapsto \heap(\ell) [\finished \mapsto \true]]$ and $\sheap' = \sheap$. Let $b = \heap(\ell)$. Since we only focus on well-formed configurations, we have $b = \obj{c}{(f \mapsto u)^*,\finished \mapsto v}$ for some activity class $c$ and some boolean value $v$. Let then $b' = \heap'(\ell) = \obj{c}{(f \mapsto u)^*,\finished \mapsto \true}$ according to the reduction rule. 

Given that $\absprog' :> \rfconf(\actconf{\actstack}{\heap}{\sheap})$, we have $\absprog' :> \rfheap(\heap)$, which implies that there exists $\absheap(\absloc,\absblock) \in \absprog'$ such that $\absloc = \rfloc(\ell)$ and $\absblock \sqsupseteq \rfblock(b)$. This means that $\absblock = \absobj{c}{(f \mapsto \absual)^*,\finished \mapsto \absval}$ for some $u^*,v$ such that $\forall i: \absual_i \sqsupseteq \rfval(u)$ and $\absval \sqsupseteq \rfval(v)$. We then observe that:
\[
\rfblock(b') = \absobj{c}{(f \mapsto \rfval(u))^*, \finished \mapsto \widehat{\true}}
\]

Since $\translate{P} \cup \rfconf(\Psi) \vdash \absprog'$ and $\absheap(\absloc,\absblock) \in \absprog'$, we have in particular $\translate{P} \cup \rfconf(\Psi) \vdash \absheap(\absloc,\absblock)$, hence:
\[
\translate{P} \cup \rfconf(\Psi) \vdash \absheap(\absloc,\absobj{c}{(f \mapsto \absual)^*,\finished \mapsto \top_{\type{bool}}}),
\]
by using the implication $\rulename{Fin}$ included in $\translate{P}$. We then observe that:
\[
\begin{array}{lcll}
\absheap(\absloc,\absobj{c}{(f \mapsto \absual)^*,\finished \mapsto \top_{\type{bool}}}) & \sqsupseteq & \absheap(\absloc,\absobj{c}{(f \mapsto \absual)^*,\finished \mapsto \widehat{\true}}) \\
& = & \absheap(\rfloc(\ell),\absobj{c}{(f \mapsto \absual)^*,\finished \mapsto \widehat{\true}}) \\
& \sqsupseteq & \absheap(\rfloc(\ell), \rfblock(b'))
\end{array}
\]
Hence, $\translate{P} \cup \rfconf(\Psi) \vdash \absprog' \cup \{\absheap(\absloc,\absobj{c}{(f \mapsto \absual)^*,\finished \mapsto \top_{\type{bool}}})\} :> \rfheap(\heap')$, which is enough to conclude the case;

\item[\irule{A-Replace}]: let $\actstack = \actframe{\ell}{\actstate{onDestroy}}{\pi}{\ocallstack} :: \actstack_0$ and $\actstack' = \uactframe{\pointer{p}{c}}{\actstate{constructor}}{\pi}{\getcb{\pointer{p}{c}}{\actstate{constructor}}} :: \actstack_0$ with $\heap(\ell) = \obj{c}{(f \mapsto v)^*,\finished \mapsto u}$, $\heap' = \heap, \pointer{p}{c} \mapsto o$ with $o = \obj{c}{(f \mapsto \defvalue_{\tau})^*,\finished \mapsto \false}$, and $\sheap' = \sheap$. Since we only focus on well-formed configurations, we know that $c$ is an activity class and $\ell = \pointer{p'}{c}$ for some pointer $p'$.

Given that $\translate{P} \cup \rfconf(\Psi) \vdash \absprog':> \rfconf(\actconf{\actstack}{\heap}{\sheap})$, we have:
\begin{itemize}
\item[(1)] $\absprog' :> \rfdispatch{\ell}(\pi)$
\item[(2)] $\absprog' :> \rfastk(\actstack_0)$
\end{itemize}
Since $\rfloc(\ell) = \rfloc(\pointer{p'}{c}) = \rfloc(\pointer{p}{c})$, from (1) we get:
\begin{itemize}
\item[(3)] $\absprog' :> \rfdispatch{\pointer{p}{c}}(\pi)$
\end{itemize}
We then observe that $\getcb{\pointer{p}{c}}{\actstate{constructor}} = \locstate{c',m,0}{\stm^*}{R}{v^*} :: \varepsilon$, where $(c',\stm^*) = \lookup(c,\actstate{constructor})$, $\sign(c',\actstate{constructor}) = \methsign{\tau_1,\ldots,\tau_n}{\tau}{\loc}$ and:
\[ 
R = ((r_i \mapsto \defvalue)^{i \leq \loc}, r_{\loc+1} \mapsto \pointer{p}{c}, (r_{\loc+1+j} \mapsto v_j')^{j \leq n}),
\]
for some values $v_1',\ldots,v_n'$ of the correct type $\tau_1,\ldots,\tau_n$. By Assumption~\ref{asm:overriding}, we also have $c \subtype c'$.

Given that $\absprog' :> \rfconf(\actconf{\actstack}{\heap}{\sheap})$, we have $\absprog' :> \rfheap(\heap)$, which implies that there exists $\absheap(\absloc,\absblock) \in \absprog'$ such that $\absloc = \rfloc(\ell) = c$ and $\absblock \sqsupseteq \rfblock(\heap(\ell))$. This implies that $\absblock = \absobj{c}{(f \mapsto \absval)^*,\finished \mapsto \absual}$ for some $\absval^*,\absual$ such that $\forall i: \absval_i \sqsupseteq \rfval(v_i)$ and $\absual \sqsupseteq \rfval(u)$. Since $\translate{P} \cup \rfconf(\Psi) \vdash \absprog'$ and $\absheap(\absloc,\absblock) \in \absprog'$, we have in particular $\translate{P} \cup \rfconf(\Psi) \vdash \absheap(\absloc,\absblock) = \absheap(c, \absobj{c}{(f \mapsto \absval)^*,\finished \mapsto \absual})$, hence:
\[
\translate{P} \cup \rfconf(\Psi) \vdash \absreg{c',m,0}{(\top_{\tau_j})^{j \leq n}}{(\adefvalue)^{k \leq \loc},c,(\top_{\tau_j})^{j \leq n}},
\]
by using the implications $\rulename{Cbk}$ included in $\translate{P}$. We then observe that:
\[
\{\absreg{c',m,0}{(\top_{\tau_j})^{j \leq n}}{(\adefvalue)^{k \leq \loc},c,(\top_{\tau_j})^{j \leq n}}\} :> \rfcall(\getcb{\pointer{p}{c}}{\actstate{constructor}})
\]
By combining (2), (3) and the last observation through Lemma~\ref{lem:join-order} we then get:
\[
\{\absreg{c',m,0}{(\top_{\tau_j})^{j \leq n}}{(\adefvalue)^{k \leq \loc},c,(\top_{\tau_j})^{j \leq n}}\} \cup \absprog' :> \rfcall(\getcb{\pointer{p}{c}}{\actstate{constructor}}) \cup \rfastk(\actstack_0) \cup \rfdispatch{\pointer{p}{c}}(\pi) = \rfastk(\actstack')
\]
Since $\translate{P} \cup \rfconf(\Psi) \vdash \{\absreg{c',m,0}{(\top_{\tau_j})^{j \leq n}}{(\adefvalue)^{k \leq \loc},c,(\top_{\tau_j})^{j \leq n}}\} \cup \absprog'$, we proved that the change to the activity stack is correctly over-approximated.

To conclude, we need to deal with the change to the heap. We first observe that $\translate{P} \cup \rfconf(\Psi) \vdash \absprog' :> \rfconf(\actconf{\actstack}{\heap}{\sheap})$ and $\rfconf(\actconf{\actstack}{\heap}{\sheap}) :> \rfheap(\heap)$, hence:
\begin{itemize}
\item[(4)] $\absprog':> \rfheap(\heap)$
\end{itemize}
Since $\translate{P} \cup \rfconf(\Psi) \vdash \absheap(\absloc,\absblock) = \absheap(c, \absobj{c}{(f \mapsto \absval)^*,\finished \mapsto \absual})$, we have\footnote{We assume here that boolean fields are initialized to $\false$. The proof can be adapted to the case where they are initialized to $\true$ by using the implication in rule $\rulename{Fin}$.}:
\[
\translate{P} \cup \rfconf(\Psi) \vdash \absheap(c,\absobj{c}{(f \mapsto \adefvalue_{\tau})^*,\finished \mapsto \widehat{\false}})),
\]
by using the implication $\rulename{Rep}$. We then observe that:
\[
\{\absheap(c,\absobj{c}{(f \mapsto \adefvalue_{\tau})^*,\finished \mapsto \widehat{\false}}))\} :> \rfheap(\pointer{p}{c} \mapsto o).
\]
By combining (4) with the latter observation by Lemma~\ref{lem:join-order}, we get:
\[
\absprog' \cup \{\absheap(c,\absobj{c}{(f \mapsto \adefvalue_{\tau})^*,\finished \mapsto \widehat{\false}}))\} :> \rfheap(\heap')
\]
Since $\translate{P} \cup \rfconf(\Psi) \vdash \absprog' \cup \{\absheap(c,\absobj{c}{(f \mapsto \adefvalue_{\tau})^*,\finished \mapsto \widehat{\false}}))\}$, we proved that also the change to the heap is over-approximated correctly;

\item[\irule{A-Hidden}]: analogous to case \irule{A-Step};

\item[\irule{A-Start}]: let $\actstack = \actframe{\ell}{s}{i :: \pi}{\ocallstack} :: \actstack_0$ and $\actstack' = \uactframe{\pointer{p}{c}}{\actstate{constructor}}{\varepsilon}{\getcb{\pointer{p}{c}}{\actstate{constructor}}} :: \actframe{\ell}{s}{\pi}{\ocallstack} :: \actstack_0$ with $i = \intent{c}{(k \mapsto v)^*}$. Also, let $\sheap' = \sheap$ and $\heap' = \heap,\heap'',\pointer{p}{c} \mapsto o, \pointer{p'}{\astart{c}} \mapsto i'$ with $\serblock{\heap}(i) = (i',\heap'')$ and $o = \obj{c}{(f \mapsto \defvalue_{\tau})^*,\finished \mapsto \false, \fintent \mapsto \pointer{p'}{\astart{c}}, \parent \mapsto \ell}$. Since we only focus on well-formed configurations, we know that $\ell = \pointer{p'}{c'}$ for some pointer $p'$ and some activity class $c'$.

Given that $\translate{P} \cup \rfconf(\Psi) \vdash \absprog':> \rfconf(\actconf{\actstack}{\heap}{\sheap})$, we have $\absprog' :> \rfdispatch{\ell}(i :: \pi)$, which implies that there exists $\absdispatch(\ann,\absblock) \in \absprog'$ such that $\ann = \rfloc(\ell) = c'$ and $\absblock \sqsupseteq \rfblock(i)$. This implies that $\absblock = \absintent{c}{\absval}$ for some $\absval$ such that $\absval \sqsupseteq \sqcup_i\, \rfval(v_i)$. We then have:
\[
\translate{P} \cup \rfconf(\Psi) \vdash \absheap(\astart{c},\absintent{c}{\absval}),
\]
and:
\[
\translate{P} \cup \rfconf(\Psi) \vdash \absheap(c, \absobj{c}{(f \mapsto \adefvalue_{\tau})^*, \finished \mapsto \widehat{\false}, \parent \mapsto c', \fintent \mapsto \astart{c}}),
\]
by using the implications $\rulename{Act}$ included in $\translate{P}$. Using the latter fact and the implications $\rulename{Cbk}$, we can prove that the change to the activity stack is over-approximated correctly, similarly to what we did in case \irule{A-Replace}: we omit details.

We focus instead on the changes to the heap. Since $\absprog' :> \rfheap(\heap)$ and $\serblock{\heap}(i) = (i',\heap'')$, we know that $\absprog' :> \rfheap(\heap'')$ by Lemma~\ref{lem:heap-serialization}. We then observe that:
\[
\{\absheap(c, \absobj{c}{(f \mapsto \adefvalue_{\tau})^*, \finished \mapsto \widehat{\false}, \parent \mapsto c', \fintent \mapsto \astart{c}})\} = \rfheap(\pointer{p}{c} \mapsto o)
\]
Finally, we notice that:
\[
\begin{array}{lcll}
\{\absheap(\astart{c},\absintent{c}{\absval})\} & :> & \{\absheap(\astart{c},\rfblock(i)\} & \text{since } \absblock = \absintent{c}{\absval}) \sqsupseteq \rfblock(i) \\
& = & \rfheap(\pointer{p'}{\astart{c}} \mapsto i) & \text{by definition} \\
& = & \rfheap(\pointer{p'}{\astart{c}} \mapsto i') & \text{by Lemma~\ref{lem:serialization}}
\end{array}
\]
By combining all these observations, we prove that the new heap is over-approximated correctly;

\item[\irule{A-Swap}]: in this case $\rfconf(\actconf{\actstack}{\heap}{\sheap}) = \rfconf(\actconf{\actstack'}{\heap'}{\sheap'})$, hence the conclusion immediately follows by the induction hypothesis;

\item[\irule{A-Result}]: let:
\[
\actstack = \actframe{\ell'}{\actstate{onPause}}{\varepsilon}{\ocallstack'} :: \actframe{\ell}{s}{\varepsilon}{\ocallstack} :: \actstack_0,
\]
and: 
\[
\actstack' = \uactframe{\ell}{s}{\varepsilon}{\getcb{\ell}{\actstate{onActivityResult}}} :: \actframe{\ell'}{\actstate{onPause}}{\varepsilon}{\ocallstack'} :: \actstack_0,
\]
with $\heap(\ell').\parent = \ell$. Also, let $\sheap' = \sheap$ and $\heap' = (\heap,\heap'')[\ell \mapsto \heap(\ell)[\result \mapsto \ell'']]$ with: 
\[
\serval{\heap}(\heap(\ell').\result) = (\ell'',\heap'').
\]
Since we focus only on well-formed configurations, we have $\ell = \pointer{p}{c}$ and $\ell' = \pointer{p'}{c'}$ for some pointers $p,p'$ and some activity classes $c,c'$. Also, let $\heap(\ell) = \obj{c}{(f \mapsto \absval)^*}$ and $\heap(\ell') = \obj{c'}{(f' \mapsto \absval')^*, \parent \mapsto \ell}$. Since $\heap(\ell) = \obj{c}{(f \mapsto \absval)^*}$, to prove that the changes to the activity stack are correctly over-approximated we can proceed like in case \irule{A-Step}, using the implications in \rulename{Cbk}: we omit details.

We focus instead on the changes to the heap. Since $\absprog' :> \rfconf(\actconf{\actstack}{\heap}{\sheap})$, we have in particular:
\begin{itemize}
\item[(1)] $\absprog' :> \rfheap(\heap)$
\end{itemize}
By (1) and $\serval{\heap}(\heap(\ell').\result) = (\ell'',\heap'')$, using Lemma~\ref{lem:heap-serialization}, we prove:
\begin{itemize}
\item[(2)] $\absprog' :> \rfheap(\heap'')$ 
\end{itemize}
Again by (1), there exists $\absheap(\absloc,\absblock) \in \absprog'$ such that $\absloc = \rfloc(\ell) = c$ and $\absblock \sqsupseteq \rfblock(\heap(\ell))$. This implies that $\absblock = \absobj{c}{(f \mapsto \absval)^*}$ for some $\absval^*$ s.t. $\forall i: \absval_i \sqsupseteq \rfval(v_i)$. Similarly, we show that there exists $\absheap(\absloc',\absblock') \in \absprog'$ s.t. $\absloc' = \rfloc(\ell') = c'$ and $\absblock' \sqsupseteq \rfblock(\heap(\ell'))$, and $\absblock' = \absobj{c'}{(f' \mapsto \absval')^*,\parent \mapsto c,\result \mapsto \ann''}$ for some $\absval'^*,\ann''$ such that $\forall i: \absval_i' \sqsupseteq \rfval(v_i')$ and $\ann'' = \rfloc(\heap(\ell').\result)$. Hence, we have:
\[
\translate{P} \cup \rfconf(\Psi) \vdash \absheap(c,\absobj{c}{(f \mapsto \absval)^*}) \wedge \absheap(c',\absobj{c'}{(f' \mapsto \absval')^*,\parent \mapsto c}),
\]
which allows us to prove:
\[
\translate{P} \cup \rfconf(\Psi) \vdash \absheap(c,\absobj{c}{(f \mapsto \absval)^*[\result \mapsto \ann'']}),
\]
by using the implication $\rulename{Res}$. We then observe that:
\[
\begin{array}{lcll}
\{\absheap(c,\absobj{c}{(f \mapsto \absval)^*[\result \mapsto \ann'']})\} & :> & \rfheap(\ell \mapsto \heap(\ell)[\result \mapsto \heap(\ell').\result]) & \text{by definition} \\
& = & \rfheap(\ell \mapsto \heap(\ell)[\result \mapsto \ell'']) & \text{by Lemma~\ref{lem:serialization}}
\end{array}
\]
Since $\heap' = (\heap,\heap'')[\ell \mapsto \heap(\ell)[\result
\mapsto \ell'']] = \heap[\ell \mapsto \heap(\ell)[\result \mapsto
\ell'']],\heap''$, by combining (1), (2) and the last observation
using Lemma~\ref{lem:join-order}, we conclude as follows:
\[
\translate{P} \cup \rfconf(\Psi) \vdash \absprog' \cup \{\absheap(c,\absobj{c}{(f \mapsto \absval)^*[\result \mapsto \ann'']})\} :> \rfheap(\heap')
\]
\end{itemize}
\end{IEEEproof}


\fi

\end{document}